\documentclass[cernpreprint, atlasdraft=false, texlive=2023, UKenglish, texmf, orcidlogo]{atlasdoc}
 
\usepackage[block=none]{atlaspackage}
\usepackage{atlasbiblatex}
 
\usepackage{atlasphysics}
 
\graphicspath{{logos/}{figures/}}
 
\usepackage{ANA-EXOT-2018-58-PAPER-defs}
\usepackage[super]{nth}
\usepackage{multirow,makecell}

\AtlasTitle{Search for pair-production of vector-like quarks in $pp$ collision
events at $\sqrt{s}=13$~TeV with at least one leptonically decaying
$Z$~boson and a third-generation quark with the ATLAS detector}
 
\PreprintIdNumber{CERN-EP-2021-207}
\AtlasAbstract{
A search for the pair-production of vector-like quarks optimized for decays into a $Z$ boson
and a third-generation Standard Model quark is presented, using the full Run~2 dataset
corresponding to $139\,\mathrm{fb}^{-1}$ of $pp$
collisions at $\sqrt{s}=13\,\mathrm{TeV}$, collected in 2015--2018 with
the ATLAS detector at the Large Hadron Collider.
The targeted final state is characterized by the presence of a $Z$ boson with high transverse momentum,
reconstructed from a pair of same-flavour leptons with opposite-sign charges, as well as by the presence
of $b$-tagged jets and high-transverse-momentum large-radius jets reconstructed from
calibrated smaller-radius jets. Events with exactly two or at least three leptons are used,
which are further categorized by the presence of boosted $W$, $Z$, and Higgs bosons and
top quarks. The categorization
is performed using a neural-network-based boosted object tagger to enhance the sensitivity to signal relative to the background.
No significant excess above the background expectation is observed and
exclusion limits at 95\% confidence level are set on the masses of the
vector-like partners $T$ and $B$
of the top  and bottom quarks, respectively.
The limits depend on the branching ratio configurations and, in the case of 100\% branching ratio
for $T\rightarrow Zt$ and 100\% branching ratio for $B\rightarrow Zb$,
this search sets the most stringent limits to date,
allowing $m_T > 1.60\,\mathrm{TeV}$ and $m_B > 1.42\,\mathrm{TeV}$, respectively.}
 
\author{The ATLAS Collaboration}
 
\AtlasRefCode{EXOT-2018-58}

\AtlasJournal{Phys. Lett. B.}
\AtlasJournalRef{Phys. Lett. B 843 (2023) 138019}
\AtlasDOI{10.1016/j.physletb.2023.138019}

\hypersetup{pdftitle={ATLAS document},pdfauthor={The ATLAS Collaboration}}
 
\begin{document}
 
\maketitle
 
\tableofcontents

\section{Introduction}
 
The Standard Model (SM) has had astonishing success in describing the
interactions of elementary particles, culminating in the
discovery~\cite{HIGG-2012-27,CMS-HIG-12-028} of the Higgs boson
by the ATLAS and CMS collaborations.
Nevertheless, there remain some shortcomings,
such as the quadratically divergent corrections predicted to contribute to the square of
the Higgs boson mass~\cite{Naturalness}.
In many theories beyond the SM, including
Composite Higgs~\cite{CompHiggs1,CompHiggs2} and
Little Higgs~\cite{LittleHiggs,LittleHiggsRev} models,
a recurrent theme that addresses these issues
is the existence of vector-like quarks (VLQs):
coloured spin-1/2 fermions that have left- and right-handed components transforming
identically under the SM gauge group. In such models,
VLQs are often assumed to couple preferentially to a third-generation
quark~\cite{VLQmixing,delAguila,delAguila:2000rc,JA_TP}, potentially regulating
the divergent corrections to the Higgs boson mass.
The VLQs therefore decay into SM quarks
and a $W/Z$ ($V$) or Higgs ($H$) boson, with branching ratios (\br)
that depend on the VLQ masses and their configuration in weak-isospin multiplets.
In renormalizable extensions of the SM that include VLQs, the canonical representation of VLQs
constitutes one of seven multiplets:
two singlets, three doublets, and two triplets.
Vector-like partners, $T$ and $B$, of the top and bottom quarks
can exist with electric charges $(2/3)e$ and $-(1/3)e$, respectively,
and can be arranged in singlets, doublets, and triplets.
VLQs with exotic charges can also exist, namely $X$ and $Y$ with
electric charges $(5/3)e$ and $-(4/3)e$, respectively, which
can be arranged in doublets and triplets with the $T$ and $B$.
Assuming an almost degenerate VLQ mass hierarchy~\cite{JA_TP},
the singlet and triplet representations are
phenomenologically similar in terms of chiral structure and BR composition.
As a consequence, the singlet and doublet representations are those that are primarily assumed in searches at the Large Hadron Collider (LHC).
While models with extreme values of 100\% BR for decay into a
third-generation quark and either a $W/Z$ boson or a Higgs boson are often used as benchmarks,
intermediate \br configurations are expected in the more physically motivated
singlet and doublet models.
In the former, the \br for $T\rightarrow Zt$ or $B\rightarrow Zb$ reaches ${\approx}25\%$ at high VLQ masses~\cite{JA_TP}.
For the $(T,B)$ and $(X,T)$ doublets the \br for $T\rightarrow Zt$ reaches ${\approx}50\%$~\cite{JA_TP}, as does
the \br for $B\rightarrow Zb$ for the $(B,Y)$ doublet.

\begin{figure}[!ht]
\centering
\includegraphics[width=.49\textwidth]{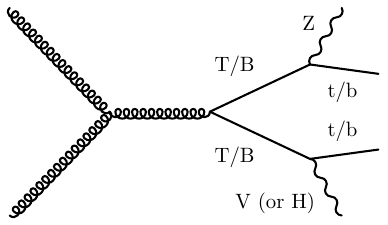}
\caption{Feynman diagram showing the pair production of $T$ or $B$ quarks in which at least
one of the VLQs decays into a $Z$ boson in association with a SM $t$- or $b$-quark, respectively.
In this analysis, events in which the $Z$ boson decays into a pair of
electrons or a pair of muons are targeted. At least one $b$-tagged jet is required to be present after
pair production of either type of VLQ, and additional
leptons can be produced in the decays of $V$ ($W$/$Z$), or $H$ bosons
on the second leg of the VLQ pair production.
}
\label{fig:feynqq}
\end{figure}
 
VLQs were searched for at ATLAS and CMS focusing mainly on the
pair-production mode~\cite{EXOT-2017-17,EXOT-2016-35,EXOT-2017-14,EXOT-2016-16,
EXOT-2017-34,EXOT-2016-13,EXOT-2016-14,EXOT-2016-15,CMS-B2G-18-005,CMS-B2G-17-012,
CMS-B2G-17-011,CMS-B2G-17-003,CMS-B2G-19-005,CMS-B2G-20-011}.  Constraints on VLQ
production were also derived recently~\cite{Buckley:2020wzk} from a range of
differential cross-section measurements at the LHC, complementing the direct searches.
VLQ pair production, proceeding primarily via
the strong interaction with a cross-section that depends only on the VLQ mass, is
expected to be the dominant mode for masses up to approximately \SI{1}{\TeV}.
The most stringent limits at 95\% confidence level (CL) on  $T$ and  $B$ masses
depend on the assumed \br configuration;
for 100\% \br for $T\rightarrow Zt$, $T\rightarrow Ht$, $T\rightarrow Wb$, and $B\rightarrow Wt$,
masses up to $\SI{1.48}{\TeV}$, $\SI{1.50}{\TeV}$,
$\SI{1.50}{\TeV}$, and $\SI{1.56}{\TeV}$ are excluded,
respectively~\cite{CMS-B2G-20-011}.
For 100\% \br for $B\rightarrow Zb$ and $B\rightarrow Hb$, $B$ masses up to
$\SI{1.39}{\TeV}$ and $\SI{1.57}{\TeV}$ are excluded, respectively~\cite{CMS-B2G-18-005}.
A singlet $T$ is excluded for masses below $\SI{1.48}{\TeV}$ and a singlet $B$ is excluded
for masses below $\SI{1.47}{\TeV}$, while in the $(T,B)$ doublet case, $T$ and $B$ masses below
$\SI{1.49}{\TeV}$ and $\SI{1.37}{\TeV}$ are excluded, respectively~\cite{CMS-B2G-20-011,EXOT-2017-17,Buckley:2020wzk}.
Single VLQ production has also been searched
for~\cite{EXOT-2016-35,EXOT-2017-16,EXOT-2016-17,EXOT-2016-16,CMS-B2G-18-003,CMS-B2G-17-018,CMS-B2G-17-009,CMS-B2G-19-004,CMS-B2G-20-002},
but the interpretation~\cite{Roy:2020fqf} of the search results
depends on an additional coupling constant for the coupling to electroweak bosons.
 
This Letter presents a search for pair production of $T$ and $B$
in events with at least two electrons or
muons where at least two same-flavour leptons with opposite-sign charges
originate from the decay of a $Z$ boson.
Various \br configurations for the decay of the VLQ into a $V$ or $H$ boson and
a third-generation quark are considered, including the singlet and doublet models, assuming
that only one type of VLQ is present, and it is also assumed
that the VLQ signal kinematics are similar for different configurations.
It is further assumed that the production of VLQ pairs is dominated by the
strong interaction and that the contribution from electroweak processes is negligible.
A diagram illustrating the targeted event topologies  is shown in \Fig{\ref{fig:feynqq}}.
The search is performed across
several different event categories included in a maximum-likelihood fit,
improving on a previous
ATLAS search~\cite{EXOT-2016-35} in the same final state.
In addition to benefiting from
the larger dataset, the search is also improved by the use of a
deep neural network (DNN) to classify the jets in each event as originating
from either a $Z$ or $W$ boson, $H$ boson, or top quark.

\section{ATLAS detector}
 
ATLAS~\cite{PERF-2007-01} is a multipurpose particle detector at the Large Hadron Collider (LHC)
with a forward--backward symmetric cylindrical geometry and a near \(4\pi\) coverage in
solid angle.\footnote{ATLAS uses a right-handed coordinate system with its
origin at the nominal interaction point (IP)
in the centre of the detector and the \(z\)-axis along the beam pipe.
The \(x\)-axis points from the IP to the centre of the LHC ring,
and the \(y\)-axis points upwards.
Cylindrical coordinates \((r,\phi)\) are used in the transverse plane,
\(\phi\) being the azimuthal angle around the \(z\)-axis.
The pseudorapidity is defined in terms of the polar angle \(\theta\) as \(\eta = -\ln \tan(\theta/2)\).
Angular distance is measured in units of \(\Delta R \equiv \sqrt{(\Delta\eta)^{2} + (\Delta\phi)^{2}}\).}
It consists of an inner tracking detector (ID)~\cite{ATLAS-TDR-2010-19,PIX-2018-001} surrounded by a thin superconducting solenoid
providing a \SI{2}{\tesla} axial magnetic field, electromagnetic and hadron calorimeters, and a muon spectrometer (MS).
The ID covers the pseudorapidity range \(|\eta| < 2.5\) and
consists of silicon pixel, silicon microstrip, and transition radiation tracking detectors.
Lead/liquid-argon (LAr) sampling calorimeters provide electromagnetic (EM) energy measurements
with high granularity.
A steel/scintillator-tile hadronic calorimeter covers the central pseudorapidity range (\(|\eta| < 1.7\)).
The endcap and forward regions are instrumented with LAr calorimeters
for both EM and hadronic energy measurements up to \(|\eta| = 4.9\).
The MS surrounds the calorimeters and uses
three large air-core toroidal superconducting magnets with eight coils each.
The field integral of the toroids ranges between \num{2.0} and \SI{6.0}{\tesla\metre}
across most of the detector.
The MS includes a system of precision chambers for tracking and fast detectors for triggering~\cite{TRIG-2016-01}.
A two-level trigger system is used to select events.
The first-level trigger is implemented in hardware and uses a subset of the detector information
to accept events at a rate below \SI{100}{\kHz}.
This is followed by a software-based trigger that
reduces the accepted event rate to \SI{1}{\kHz} on average
depending on the data-taking conditions.
An extensive software suite~\cite{ATL-SOFT-PUB-2021-001} is used in the reconstruction
and analysis of real and simulated data, in detector operations, and in the trigger
and data acquisition systems of the experiment.

\section{Data and simulated event samples}
 
The analysed dataset comprises $pp$ collisions
at a centre-of-mass energy $\sqrt{s}=\SI{13}{\TeV}$
recorded by ATLAS between 2015 and 2018 with all detector
subsystems operational and with the LHC operating in
stable beam conditions with $\SI{25}{ns}$ bunch spacing.
The combined Run~2 dataset corresponds to an integrated luminosity of \lumi\
with an average of about 34 simultaneous interactions per bunch crossing (pile-up).

All the nominal Monte Carlo (MC) simulation samples used in the analysis were produced with the ATLAS
full-simulation framework~\cite{SOFT-2010-01}
based on \GEANT~\cite{Agostinelli:2002hh}.  In all samples, pile-up was
modelled by combining simulated inelastic $pp$ events with the physics event.
The nominal sample for
$Z$ boson production in association with jets (\zjets) was generated with
\SHERPA[2.2.1]~\cite{Gleisberg:2008ta, Hoeche:2009rj, Gleisberg:2008fv, Schumann:2007mg} and
the nominal diboson (\VV) sample
was generated with \SHERPA[2.2.2], with the NNPDF3.0~\cite{Ball:2014uwa} next-to-next-to-leading-order (NNLO) parton distribution function (PDF) set.
The \zjets sample includes events generated with up to two partons at  next-to-leading order
(NLO) and up to four partons at leading order (LO) and is normalized to the NNLO cross-section~\cite{Anastasiou:2003ds}.
The  \VV sample is normalized to the \SHERPA\ NLO cross-section and
includes $q\bar{q}$-initiated events with up to one parton
at NLO and up to three partons at LO and $gg$-initiated
processes generated using LO matrix elements for up to one additional jet.
For both samples, \textsc{Comix}~\cite{Gleisberg:2008fv} and \OPENLOOPS~\cite{Cascioli:2011va} were used and
the matrix element (ME) was merged with the \SHERPA\ parton shower~\cite{Schumann:2007mg} according to the \MEPSatNLO
prescription~\cite{Hoeche:2012yf}.
To estimate modelling uncertainties, an additional \zjets sample was produced with
\MGNLO[2.2.3]~\cite{Alwall:2014hca}, using the \NNPDF[3.0nlo] PDF set and interfaced to
\PYTHIA[8.210]~\cite{Sjostrand:2014zea} with the A14 set of tuned parameters (tune)~\cite{ATL-PHYS-PUB-2014-021} and the \NNPDF[2.3lo] PDF for showering.
These samples were generated including LO matrix elements for up to four jets~\cite{ATL-PHYS-PUB-2016-003}.
 
The nominal SM \ttbar background sample uses the \POWHEG
method~\cite{Nason:2004rx, Frixione:2007vw}
implemented in \POWHEGBOX~v2~\cite{Alioli:2010xd, Campbell:2014kua} with the \NNPDF[3.0nlo] PDF
set. \POWHEGBOX was interfaced with \PYTHIA[8.230]
with the A14 tune for showering.
The sample is normalized to the NNLO cross-section in QCD including resummation
of next-to-next-to-leading logarithmic (NNLL) soft gluon
terms calculated with \TOPpp~\cite{Czakon:2011xx,Czakon:2013goa, Beneke:2011mq, Cacciari:2011hy, Czakon:2012pz, Czakon:2012zr, Baernreuther:2012ws}.
For the evaluation of modelling uncertainties, samples were produced with the same ME generator as the nominal
sample, but \HERWIG[7] was used with the H7-UE-MMHT tune~\cite{Bellm:2015jjp} for the showering.
Additional samples\cite{ATL-PHYS-PUB-2018-009} were generated with \MGNLO[2.3.3] and the \NNPDF[3.0nlo] PDF set,
using the same showering configuration as the nominal sample.
 
The nominal sample including \ttbar production in
association with a vector boson (\ttbarV) was generated  with \MGNLO[2.3.3] interfaced with \PYTHIA[8.210] for
showering, using the \NNPDF[2.3lo] PDF set and the A14 tune. The sample includes  \ttbarZ and \ttbarW events normalized
to the  NLO cross-sections calculated with \MGNLO.
To evaluate modelling uncertainties, samples were produced using \SHERPA[2.2.1].
Alternative samples were also produced
with the nominal sample's ME generator, and in these samples either the A14 tune was varied or \HERWIG[7] was used with the H7-UE-MMHT tune for the showering.
The nominal \ttbarV sample also includes $t\bar{t}t\bar{t}$ and $t\bar{t}WW$ events
generated at LO and normalized to cross-sections
calculated~\cite{Alwall:2014hca,Frederix:2017wme} with NLO QCD and EW corrections.
 
The single-top processes were simulated with \POWHEGBOX~\cite{Frederix:2012dh,Re:2010bp}
using the \NNPDF[3.0nlo] PDF set and interfaced to \PYTHIA[8.234] with the A14 tune. The
samples are normalized to their respective NLO QCD cross-sections~\cite{Aliev:2010zk, Kant:2014oha} for the t-channel
and s-channel, and with additional
NNLL soft gluon terms for $Wt$ production~\cite{Kidonakis:2011wy, Kidonakis:2010ux,Kidonakis:2010tc}. The diagram-removal
scheme~\cite{Frixione:2008yi} was used in the generation of $Wt$ events to address overlaps with the \ttbar sample.

Signal samples for the pair production of $T$ and $B$ quarks were
generated with \PROTOS~\cite{PROTOS} interfaced with \PYTHIA[8.186]
using the \NNPDF[2.3lo] PDF set and the A14 tune. The samples were generated under the assumption
that the coupling is small therefore the width of the VLQ is driven by the experimental resolution.
The generated VLQ masses have widths typically smaller than 2\% of the mass.
Masses from $\SI{600}{\GeV}$ to $\SI{2}{\TeV}$ were simulated
in the singlet model, but with the \br for each of the three decays to a vector or Higgs boson
fixed to $1/3$ for all samples. A \br reweighting procedure is performed event-by-event
in order to achieve any \br configuration for a given VLQ mass, including the configurations expected
for the doublet models. Dedicated signal samples in the doublet models for the $\SI{700}{\GeV}$ and $\SI{1.2}{\TeV}$
mass points were also produced to validate the reweighting procedure and to verify
that the singlet and doublet signal kinematics are indistinguishable in the analysis as is assumed.
The signal sample cross-sections were
calculated with \TOPpp at NNLO+NNLL in QCD using the \MSTW[nnlo]~\cite{Martin:2009iq,Martin:2009bu,Martin:2010db} PDF set.

\section{Object reconstruction}

Events are required to have at least one vertex candidate with at least
two tracks with transverse momentum $\pt > \SI{0.5}{\GeV}$. The primary vertex (PV)
is defined to be the candidate with the largest $\Sigma \pt^2$,
where the sum is performed over all associated tracks.
 
Electrons are reconstructed~\cite{EGAM-2018-01}
from clusters in the EM calorimeter matched with ID tracks
and must fulfil the \textit{tight likelihood}
identification criteria. Electrons are
calibrated and are required to have
$\pt > \SI{28}{\GeV}$ and to be reconstructed  within $|\eta|<2.47$,
excluding the barrel--endcap transition regions ($1.37 < |\eta|< 1.52$).
In order to maintain a high acceptance for the expected signal events,
no isolation requirements are applied to electron candidates
beyond those implicit in the trigger requirements.
Furthermore, electron candidates must be associated with the PV
by requiring that the longitudinal impact parameter with respect to the PV
satisfies $|z_0 \cdot \sin\theta| < \SI{0.5}{mm}$ and
that the transverse impact parameter with respect to the beamline ($d_0$)
has a significance $|d_0|/\sigma(d_0) < 5$.
 
Muons are
reconstructed~\cite{MUON-2018-03} from combined tracks in the MS
and the ID and must fulfil \textit{medium}
identification criteria. Muons are calibrated~\cite{MUON-2022-01} and are
required to have $\pt>\SI{28}{\GeV}$ and to be
reconstructed  within $|\eta|<2.5$.  Muon candidates must
also satisfy the track-based isolation requirements
defined by the \textit{FixedCutTightTrackOnly} working point.
This working point uses the scalar sum of the \pt of all tracks
that are within a cone of size
$\Delta R = \min\left\{0.3,\SI{10}{\GeV}/\pt(\mu)\right\}$
around the muon candidate,
where $\pt(\mu)$ is the candidate muon $\pt$. The track associated
with the muon candidate under consideration is excluded from the sum.
The muon is selected if this sum is less than 15\% of $\pt(\mu)$.
Finally, muon candidates are required to
have $|z_0 \cdot \sin\theta| < \SI{0.5}{mm}$
and a $d_0$ significance smaller than $3$.
 
Jets are reconstructed using the \antikt\ algorithm~\cite{Cacciari:2008gp,Fastjet}
with a radius parameter of 0.4 from topological clusters of energy deposits in the
calorimeter~\cite{PERF-2014-07,ATL-PHYS-PUB-2015-036}.
Jets are calibrated to an energy scale obtained from a combination
of simulation-based corrections and measurements in data~\cite{PERF-2016-04}
and are required to fulfil $\pt > \SI{25}{\GeV}$ for
$|\eta|<2.5$ and $\pt>\SI{35}{\GeV}$ for $2.5<|\eta|<4.5$.
To reduce jet contributions from pile-up, a `jet vertex tagger' algorithm
using a two-dimensional likelihood discriminant~\cite{PERF-2014-03}
is applied to jets with $|\eta|<2.4$ and $\pt < \SI{60}{\GeV}$.
The MV2c10 algorithm~\cite{FTAG-2018-01} is used to
identify jets  in the central region ($|\eta|<2.5$)
containing a $b$-hadron decay ($b$-tagging) with a working point corresponding
to a $b$-tagging efficiency in simulated $\ttbar$ events of $77\%$,
a $c$-jet rejection factor of ${\sim}6$, and a light-jet rejection factor of ${\sim}130$.
 
The missing transverse momentum~\cite{PERF-2016-07}, with magnitude \met,
is defined as the negative vectorial
sum of the transverse momenta of all the calibrated reconstructed lepton and jet candidates in the event and includes a `soft term' with contributions
from tracks emanating from the PV but not associated with any of the reconstructed objects.
 
A procedure to remove potential overlaps between reconstructed leptons and jets is
performed sequentially as follows. First, any
muon that leaves energy deposits in the calorimeters and shares a track in the
ID with an electron is removed. After such muons have been removed,
any electron sharing an ID track with one of the
remaining muons is removed. Next, any jet within $\Delta R =0.2$
of an electron is removed, followed by the removal of electrons within $\Delta R =0.4$ of
any remaining jet.
Subsequently, any jet with at most two
tracks with $\pt > \SI{0.5}{\GeV}$ within $\Delta R= 0.2$ of a
muon is removed, unless it has been $b$-tagged.
At the end of the procedure, any muon within
$\Delta R = \min\left\{0.4,0.04 + \SI{10}{\GeV}/\pt(\mu)\right\}$ of any remaining jet is removed.

Finally, large-radius `reclustered' (RC)
jets~\cite{Nachman:2014kla} are reconstructed
by applying the \antikt\ algorithm with a radius parameter of 1.0, using the set
of selected and
already calibrated smaller-radius jets defined above.
To reduce dependence on pile-up, the RC jets are trimmed~\cite{Krohn:2009th} by
removing all constituent jets with \pT below 5\% of the RC jet \pT.
These RC jets are used
only for event categorization, and therefore no additional checks for potential overlaps
with the previously defined objects are performed.

\section{Event  selection and categorization}

An initial preselection of events is performed as follows.
Events are required to satisfy at least one of the single-lepton triggers
operating during Run~2~\cite{TRIG-2016-01,TRIG-2018-01,TRIG-2018-05}.
These triggers had varying \pt thresholds for electrons and muons
for different data-taking periods, as well as isolation requirements that
were typically more relaxed with increasing \pt threshold.
For data collected in 2015, the thresholds are 24, 60, and \SI{120}{\GeV}
for electrons and 20 and \SI{50}{\GeV} for muons. For the remainder of the dataset,
the thresholds are slightly increased to 26, 60, and \SI{140}{\GeV} for electrons
and to 26 and \SI{50}{\GeV} for muons.
Events are additionally required to have at least two
opposite-sign-charge, same-flavour (OS-SF) leptons
and to have at least two jets in the central region.
The pair of OS-SF leptons with an invariant mass \mll closest to $\mz=\SI{91.2}{GeV}$
is referred to as `the $Z$ boson candidate' and only events with $|\mll-\mz|<\SI{10}{\GeV}$
are kept.

\begin{figure}[!htbp]
\centering
\subfloat[]{
\includegraphics[width=.49\linewidth]{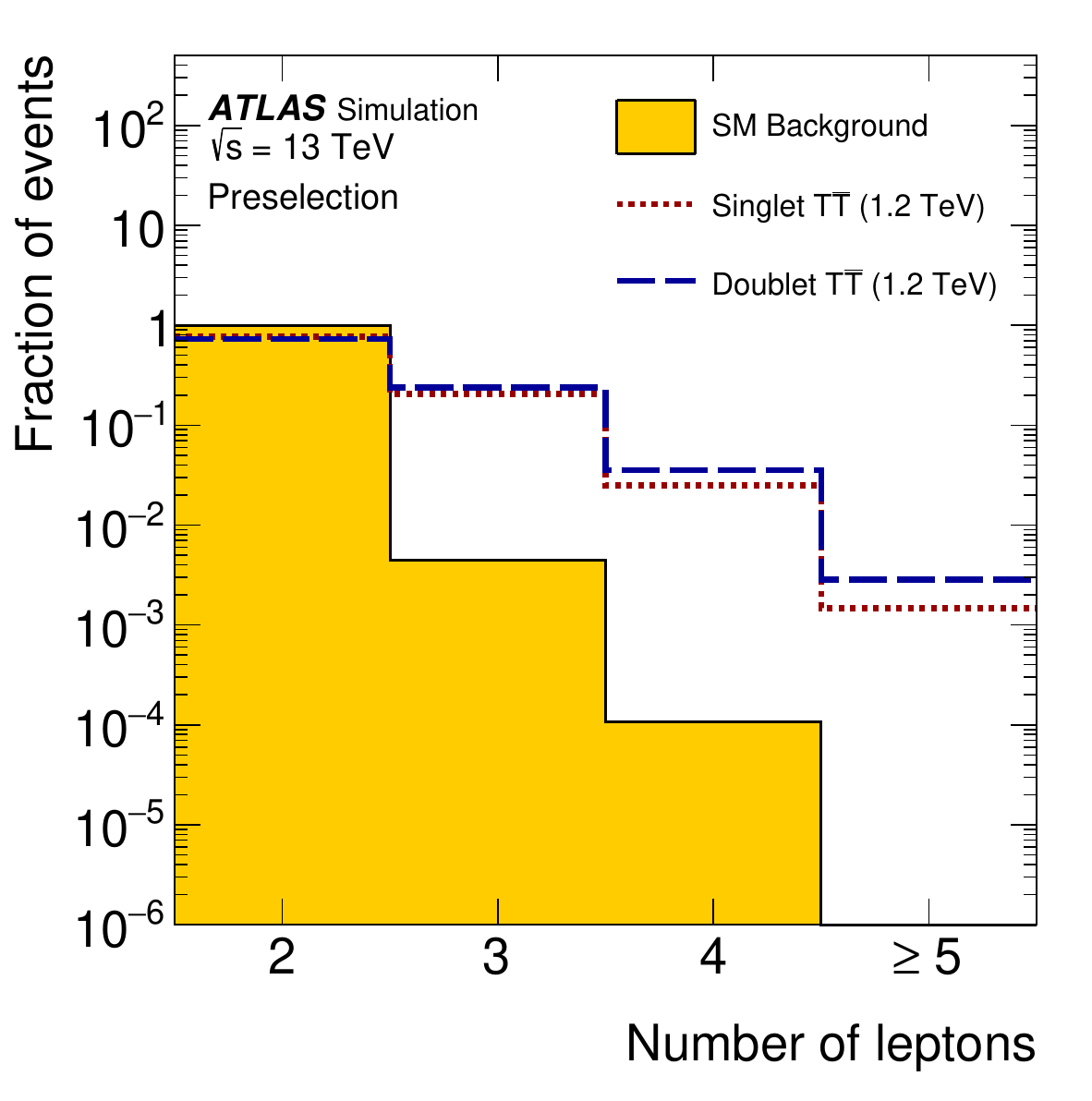}
}
\subfloat[]{
\includegraphics[width=.49\linewidth]{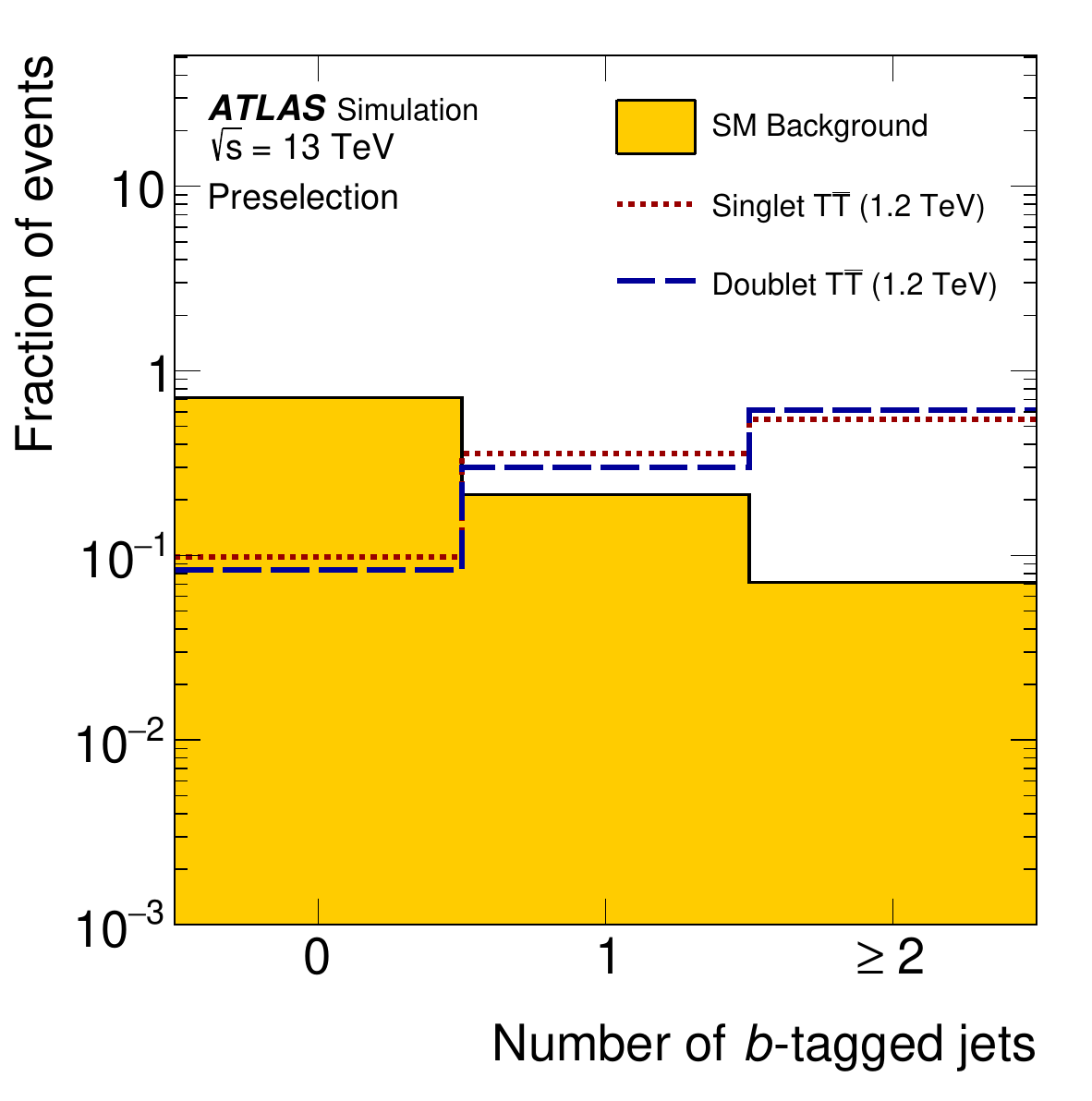}
}\\
\subfloat[]{
\includegraphics[width=.49\linewidth]{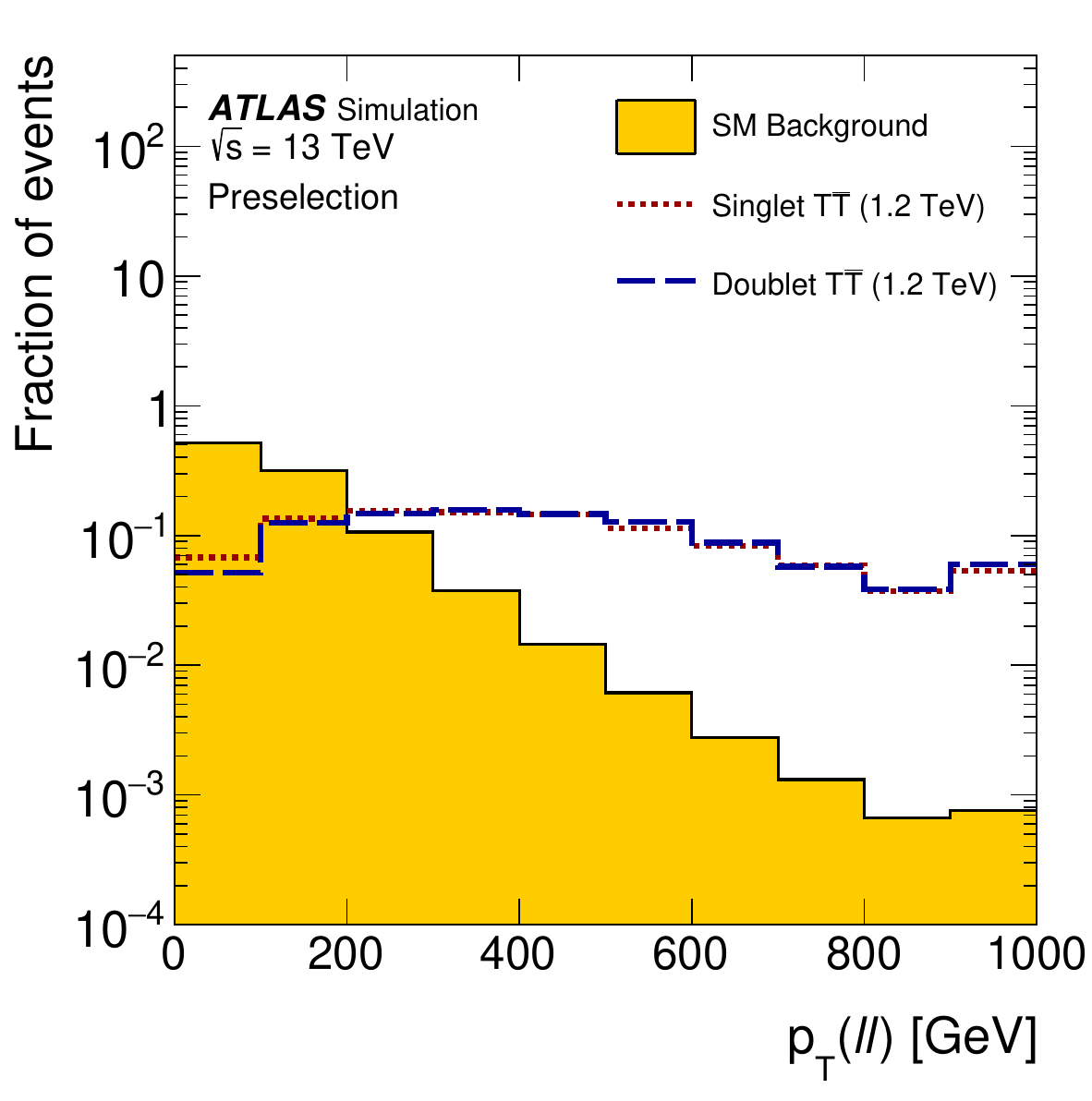}
}
\subfloat[]{
\includegraphics[width=.49\linewidth]{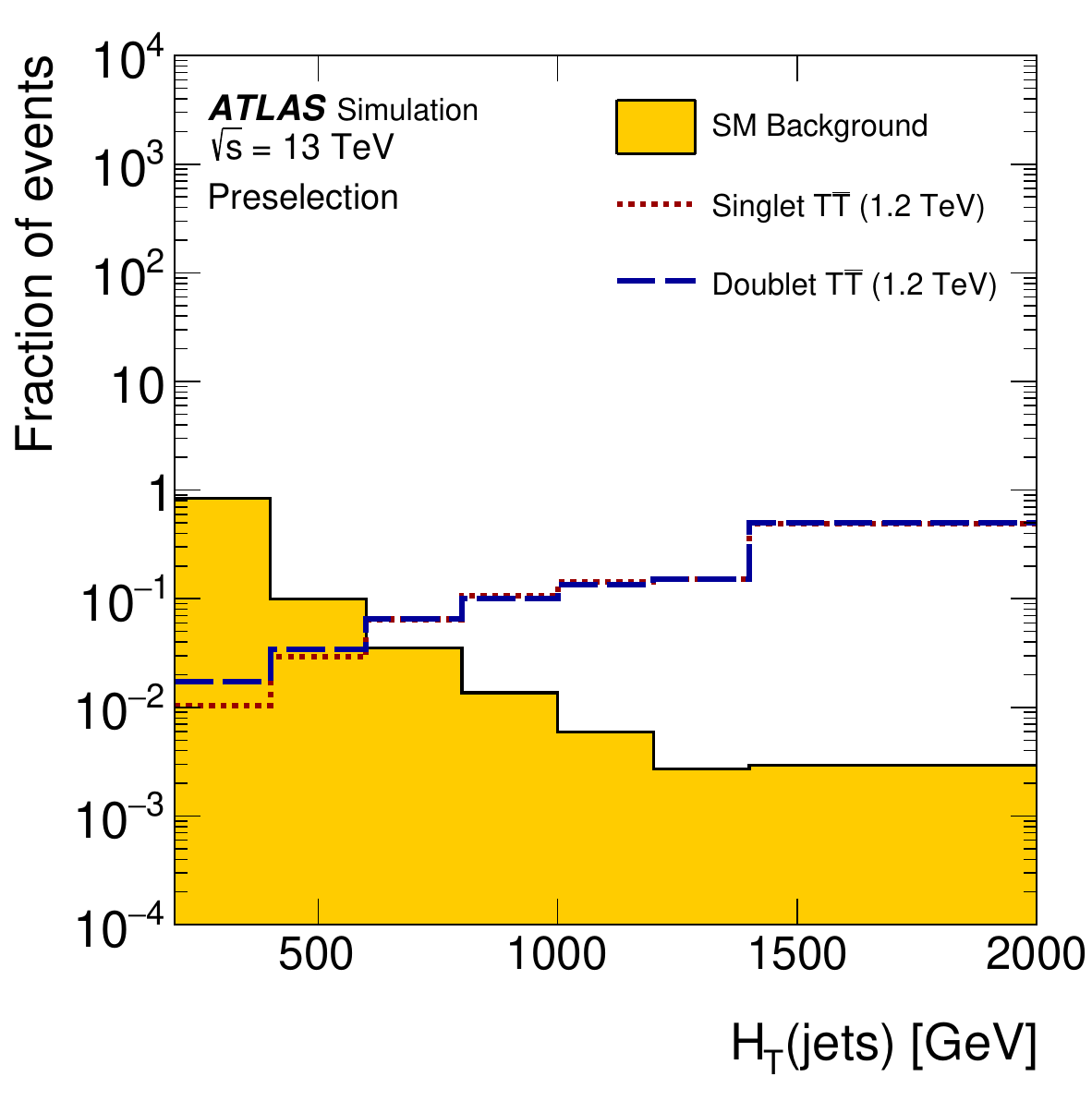}
}
 
\caption{Distributions, normalized to unit area,
obtained from MC simulation for the sum of all the
background processes (solid area) and for benchmark signal processes (dashed lines):
(a) lepton multiplicity, (b)
$b$-tagged jet multiplicity, (c) transverse momentum of the $Z$ boson candidate, \ptz, and
(d) scalar sum of the jet transverse momenta, $\htjet$. The signal processes concern the
pair production of vector-like $T$ or $B$ quarks with a mass of \SI{1.2}{\TeV} in either the
singlet or doublet configurations. The last bin contains the overflow.}
\label{fig:shapePlots}
\end{figure}

Preselected events are divided into two orthogonal and individually optimized
channels: one requiring exactly two leptons (dilepton, labelled as `\ppdil') and
a second requiring at least three leptons (trilepton, labelled as `\pptril').
The channels are combined statistically to obtain the final result,
taking advantage of the relatively high signal acceptance achieved by the \ppdil channel
and the higher signal purity offered by the \pptril channel.

\Fig{\ref{fig:shapePlots}} shows distributions of example kinematic variables
for the expected background and for benchmark signal processes obtained from MC simulation.
In addition to the lepton multiplicity, the analysis
exploits the high multiplicities of jets, large-radius jets, and $b$-tagged jets
expected for pair-produced VLQ signal.
Requirements on the momentum of the $Z$ boson candidate, \ptz,
and the scalar sums of the transverse momenta of objects reconstructed
in the events, $H_\mathrm{T}$, are applied to
suppress the background.
Furthermore,  the analysis uses \rcjets\
as an input to a `multi-class boosted object tagger' (MCBOT) in order to identify (tag)
the origin of each \rcjet\ as being either a hadronically decaying $V$ boson, $H$ boson,
or top quark.

\begin{figure}[t]
\centering
\includegraphics[width=.8\linewidth]{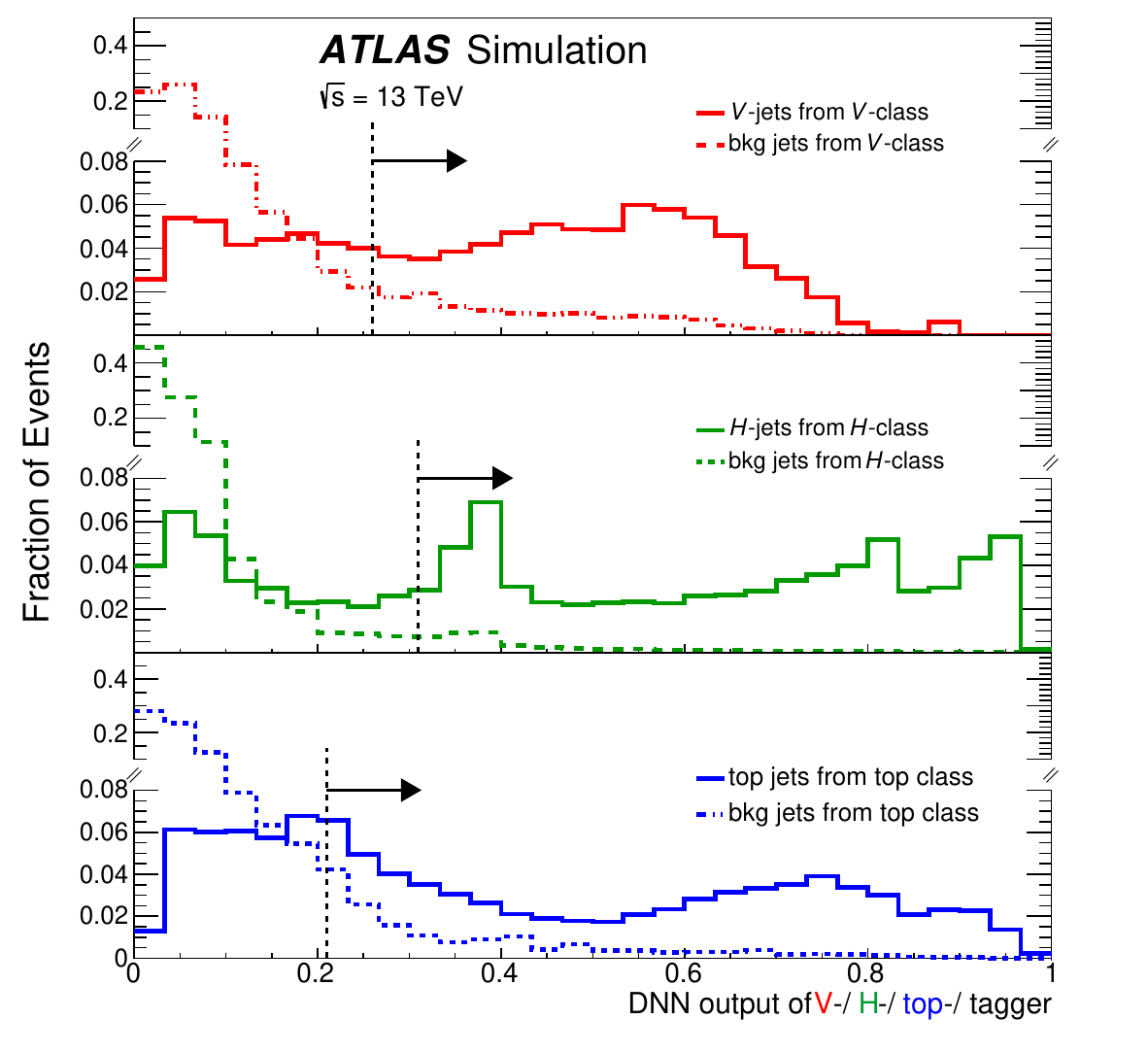}
\caption{The distributions of the DNN output of MCBOT for the $V$ boson, $H$ boson, and top
quark classes for RC jets with \pt between $\SI{150}{\GeV}$ and $\SI{1}{\TeV}$.
For each tagger ($V$, $H$, or top) the response to the corresponding signal jets
(solid lines) is compared with the response to  the background jets (dashed lines).
The responses to signal jets other than the corresponding signal
for each tagger are not shown for simplicity. Signal jets
are defined by geometrically matching the \rcjets\ to the corresponding generator-level
particles. Background jets are obtained from multijet events. The colours red, green, and blue
indicate the DNN output values for the $V$-, $H$-, or top-tagger, respectively.
The \rcjet\ \pt is reweighted together with the
\rcjet $\eta$ to be uniform over the full range. All distributions are normalized to unit area.
The $y$-axis is split for better readability.}
\label{fig:mcbotStrategy}
\end{figure}
 
MCBOT is based on a multi-class DNN using the Keras~\cite{chollet2015keras}
and TensorFlow~\cite{tensorflow2015-whitepaper}
software libraries. The DNN is trained using \rcjets\ from simulated
$Z^\prime\rightarrow\ttbar$ events, $W^\prime\rightarrow WZ$ events,
and events with a Kaluza--Klein graviton in the bulk Randall--Sundrum model~\cite{Agashe:2007zd}
decaying into a pair of $H$ bosons which each
decay into \bbbar with a \br fixed to 100\%.
These dedicated events are generated so that
\rcjets with  a mass of  at least $\SI{40}{\GeV}$
are uniformly distributed in an \rcjet \pt range between $\SI{150}{\GeV}$ and $\SI{3}{\TeV}$.
The three signal labels ($V$, $H$, top)
are assigned by matching the \rcjet\ to the corresponding
hadronically decaying boson or top quark within a $\Delta R$ of 0.75 at generator level.
Simulated multijet events are used to obtain the \rcjets\ with a \textit{background} label.
The \rcjets\ in these four classes are reweighted such that
their distributions are uniform in \rcjet \pt and $\eta$.
The 18 input variables of the DNN are the \rcjet\ \pt and mass,
the number of the \rcjet\ constituent smaller-radius jets,
the four-momentum vectors of the three highest-$\pt$ constituent jets,
and the $b$-tagging decisions for those three jets.
The DNN consists of four fully connected hidden layers with \textit{Rectified Linear Unit}~\cite{Schmidhuber_2015} activation functions
and a four-dimensional
output layer in which nodes are activated using the \textit{softmax}~\cite{Goodfellow-et-al-2016} function.
The first hidden layer uses 32 nodes, reduced to 27, 14, and 12 nodes for
the second, third, and fourth hidden layer, respectively.
The ADAM~\cite{kingma2014adam} optimizer
is used to minimize the categorical cross-entropy as a
loss function in order to find the optimal weights.
The set of hyper-parameters as optimized for a related ATLAS
analysis~\cite{EXOT-2017-14} is used.
The projections of the
four-dimensional output of the
DNN are used to define a working point for each
of the signals in order to define a $V$-tag, a $H$-tag, and a top-tag. \Fig{\ref{fig:mcbotStrategy}} shows
an example set of these projections for RC jets with \pt between $\SI{150}{\GeV}$ and $\SI{1}{\TeV}$.
In cases where the same \rcjet satisfies the requirements for more than one tag,
the tag with the highest neural-network output value is retained. Therefore, at
most one signal label can be assigned to any RC jet. For the selected working point,
and for \TTbar signal with $m_T=\SI{1400}{\GeV}$ in the singlet configuration,
efficiencies of 55\%, 41\%, and 54\% are measured for $V$, $H$, and top signals, respectively.

In the \ppdil channel, events are required to satisfy $\ptz > \SI{300}{\GeV}$
in order to select leptonic $Z$ boson candidates
originating from the decay of a heavy VLQ. In addition, given the
large amount of jet activity expected in signal events and to take advantage of a
potentially invisible decay of the $Z$ boson in the decay chain of the
second VLQ in the pair, requirements are placed on the scalar sum
of the jet transverse momenta, \htjet, and  \met.
Two exclusive signal regions (SR) are defined by requiring
$\htmet>\SI{1380}{\GeV}$ and either
exactly one \btagged jet ($1b$ SR) or at least two \btagged jets ($2b$ SR).
The chosen kinematic requirements are motivated by the selection criteria of the
earlier ATLAS analysis~\cite{EXOT-2016-35} and are optimized using a
procedure that maximizes the expected sensitivity for a few benchmark signal models.
The most relevant event selection criteria are shown in Table~\ref{tab:selection},
which additionally shows the definitions of the signal regions and background
control regions.
The events in each SR are divided
into exclusive categories based on combinations of the
numbers of signal-tagged \rcjets\ of various types
identified by MCBOT. A signal significance
optimization procedure is used to group the possible tagging combinations
so as to reduce the set of categories to seven, as shown in Table~\ref{tab:mcbot_cats}.
These include `Double tag' categories with different combinations
of at least two RC jets with tags of any signal type, as well as a final
`Overflow' category that includes
events with more than two tags but also leftover two-tag combinations
that are not included in the former categories.
In order to recover some of the events in which
RC jets might be misidentified, combinations with more than one $V$ or
$H$ boson are included even though one of the two VLQs
is required to decay to a leptonically decaying $Z$ boson
in the targeted event topology.
 
\begin{table}[t]
\caption{Definitions of analysis regions and the most relevant event selection criteria.
The final discriminating variables used
in the fit are also indicated for each region:
\mzb and \mziindb refer to the mass of the system of the $Z$ boson candidate and either the
leading $b$-jet or   subleading $b$-jet, respectively.
\htjet is defined as the scalar sum of the transverse momenta of all selected jets,
while \htjetlep is defined as the scalar sum of the transverse momenta of all
selected jets and leptons.}
\label{tab:selection}
\centering
\resizebox{\textwidth}{!}{
\begin{tabular}[ht]{lcccccccc}
\toprule{{Preselection}} &\multicolumn{8}{c}{$\geq2$ central jets} \\
&\multicolumn{8}{c}{at least two SF leptons with $\pt>\SI{28}{\GeV}$} \\
&\multicolumn{8}{c}{at least one pair of OS-SF leptons $|\mll-\mz| < \SI{10}{\GeV}$} \\
\midrule{Channel}  &\multicolumn{4}{c}{{\ppdil}} & \multicolumn{4}{|c}{{\pptril}}\\
{definitions}   &\multicolumn{4}{c}{$=2\ell$} & \multicolumn{4}{|c}{$\geq3\ell$}\\
&\multicolumn{4}{c}{$\ptz > \SI{300}{\GeV}$} &  \multicolumn{4}{|c}{$\ptz > \SI{200}{\GeV}$}\\
&\multicolumn{4}{c}{$\htjet+\MET> \SI{920}{\GeV}$} & \multicolumn{4}{|c}{$\htjetlep>\SI{300}{\GeV}$}\\
\midrule{Region}    & {$1b$  SR} & {$2b$  SR} & {$1b$  CR} &{$2b$  CR} & \multicolumn{2}{|c}{{SR}}& \multicolumn{2}{r}{{$VV$ CR}}\\
{definitions}    &\multicolumn{2}{c}{$\htjet+\MET> \SI{1380}{\GeV}$} & \multicolumn{2}{c}{$\htjet+\MET< \SI{1380}{\GeV}$} &  \multicolumn{2}{|c}{--}    &\multicolumn{2}{r}{--} \\
&{$=1~b$-jet} &{$\geq2~b$-jet} & {$=1~b$-jet} & {$\geq2~b$-jet} &  \multicolumn{2}{|c}{$\geq 1~b$-jet}    &\multicolumn{2}{r}{$=0~b$-jet} \\
 
\midrule{MCBOT categories} & 7 & 7 & -- & -- & \multicolumn{2}{|c}{5}& \multicolumn{2}{r}{--}\\
{Fitted variable} & \mzb & \mziindb&\multicolumn{2}{c}{$\htjet+\MET$} & \multicolumn{4}{|c}{\htjetlep}  \\
\bottomrule
\end{tabular}
}
\end{table}
 
In the \pptril channel, the $Z$ candidate transverse
momentum requirement is relaxed to $\ptz>\SI{200}{\GeV}$ to increase signal
efficiency given the significant background suppression due to the
three-lepton requirement.
To take advantage of the additional lepton activity, events must also
satisfy $\htjetlep>\SI{300}{\GeV}$, where \htjetlep is the scalar sum of the jet and lepton transverse momenta.
Events in the \pptril SR are also required to have at least one \btagged jet and are divided into
five exclusive categories based on the number of MCBOT tags, similarly to the \ppdil channel, as shown in Table~\ref{tab:mcbot_cats}.
 
The signal efficiency, defined as the ratio of the number of events selected by the signal regions to the number of events generated in the signal samples, for the benchmark of a $\SI{1.2}{\TeV}$ singlet $T$, is 0.3\% and 0.5\% in the 1$b$ and 2$b$ \ppdil signal regions, respectively, and 0.5\% in the \pptril signal regions.
 
The SM background contribution is estimated
using MC simulation and is adjusted in the
final simultaneous maximum-likelihood fit to the data with the help of
signal-depleted control regions (CR),  designed to be similar in phase-space to the SR.
Leading backgrounds include \zjets, which is the dominant background for \ppdil,
as well as \VV and \ttbarV, which are dominant for \pptril.
Smaller backgrounds include those from \ttbar, single top quark and four top quark.
 
The inputs to the fit are
binned distributions of final discriminants optimized for each of the two channels.
In the \ppdil channel, in all the $1b$ SR categories, the invariant mass of the leptonically decaying $Z$ candidate
and the leading \btagged jet, \mzb, is used as a final discriminant,
while for the $2b$ SR categories the subleading \btagged jet is used instead
and the discriminating variable is \mziindb.
Two \ppdil CR corresponding to each SR are defined
by requiring  $\SI{920}{\GeV}<\htmet<\SI{1380}{\GeV}$. These CR are not further categorized
using MCBOT and participate in the fit using the distribution of \htmet.
In the \pptril channel a  single \VV~CR is defined by
requiring exactly zero $b$-jets.
The final discriminant for all \pptril SR categories and for the \VV~CR is \htjetlep.

\begin{table}[t]
\caption{Definitions of the categorization of events in the \ppdil and \pptril channels
based on the number of MCBOT tags. Unless otherwise noted, the columns indicate the exact
number of required tagged RC jets for each type.
Categories labelled as `Double tag' include
combinations with exactly two tags of any type.
Categories labelled as `Overflow (OF)'
include all the tagging combinations that are
not explicitly included in the other categories.}
\label{tab:mcbot_cats}
\centering
\begin{tabular}{l ccc |ccc | ccc}
\toprule
\multirow{3}{*}{{Category}}
& \multicolumn{6}{c|}{{\ppdil channel}} & \multicolumn{3}{c}{{\pptril channel}}\\
& \multicolumn{3}{c|}{{$1b$ SR}} & \multicolumn{3}{c|}{{$2b$ SR}} & \multicolumn{3}{c}{ -- }\\\cline{2-10}
&  $V$-tags  &  $H$-tags  &  top-tags  &  $V$-tags  &  $H$-tags  &  top-tags &  $V$-tags  &  $H$-tags  &  top-tags \\
\midrule{No tag } & 0 & 0 & 0         & 0 & 0 & 0        & ~\,0 & ~\,0 & ~\,0        \\\midrule{V tag}   & 1 & 0 & 0         & 1 & 0 & 0        & $\geq 1$~~ & ~\,0 & ~\,0 \\\midrule{H tag }  & 0 & 1 & 0         & 0 & 1 & 0        & ~\,0 & $\geq 1$~~ & ~\,0 \\\midrule{Top tag} & 0 & 0 & 1         & 0 & 0 & 1        & ~\,0 & ~\,0 & $\geq 1$~~ \\\midrule
 
\multirow{4}{*}{{Double tag 1}} & 2 & 0 & 0      & 2 & 0 & 0 &        \multicolumn{3}{c}{--} \\
& 0 & 2 & 0      & 0 & 2 & 0 &        \multicolumn{3}{c}{--} \\
& 1 & 0 & 1      & 1 & 1 & 0 &        \multicolumn{3}{c}{--} \\
&\multicolumn{3}{c|}{--}      & 0 & 0 & 2 &        \multicolumn{3}{c}{--} \\\midrule
 
\multirow{2}{*}{{Double tag 2}} & 0 & 1 & 1      & 0 & 1 & 1 &        \multicolumn{3}{c}{--} \\
& 0 & 0 & 2      & \multicolumn{3}{c|}{--} &        \multicolumn{3}{c}{--} \\\midrule
 
\multirow{4}{*}{{Overflow (OF)}}  & 1 & 1 & 0    & 1 & 0 & 1                     & ~\,0        & $\geq 1$~~ & $\geq 1$~~ \\
& \multicolumn{3}{c| }{or $>2$ tags} &  \multicolumn{3}{c|}{or $>2$ tags}    & $\geq 1$~~ & ~\,0  & $\geq 1$~~ \\
& \multicolumn{3}{c| }{--} &  \multicolumn{3}{c|}{--}    & $\geq 1$~~ & $\geq 1$~~ & ~\,0        \\
& \multicolumn{3}{c| }{--} &  \multicolumn{3}{c|}{--}    & $\geq 1$~~ & $\geq 1$~~ & $\geq 1$~~ \\\bottomrule
 
\end{tabular}
\end{table}

\section{Systematic uncertainties}
 
Several experimental and theoretical systematic uncertainties that can
affect the normalization or the shape of the fitted distributions are considered.
For each considered  source of uncertainty,  variations representing
the $-1\sigma$ and $+1\sigma$ confidence interval are derived.
The analysis selection,  including the MCBOT categorization, is applied on each
variation to estimate their effect on the final discriminants.
Several of these modelling uncertainties are constrained after the fit
due to the significantly larger number of events in the control regions
compared to the signal regions.  However, these constraints do not impact the final limit.
Furthermore,  the analysis is dominated by statistical uncertainties;
the expected lower limits on the VLQ masses decrease by no more than 4\%
when all the systematic uncertainties are included in the analysis.

Experimental uncertainties include effects on the electron energy scale and energy resolution~\cite{EGAM-2018-01},
the muon momentum scale and resolution~\cite{MUON-2018-03}, as well as uncertainties in the data-to-MC correction
factors for the electron and muon trigger, reconstruction, identification, and isolation
efficiencies~\cite{EGAM-2018-01,MUON-2018-03}. Jet energy scale and resolution uncertainties are also included,
as obtained from studies in data and simulation~\cite{PERF-2016-04}.
A 10\% uncertainty~\cite{JETM-2018-06}  is assigned to the small-radius jet mass, which is
not calibrated but is used in the \rcjet\ mass calculation.
Flavour-tagging
uncertainties include uncertainties in the \bjet tagging, \cjet tagging, and light-jet
mis-tagging efficiencies, and uncertainties due to extrapolations to regions not covered by
the data used for the efficiency measurements~\cite{FTAG-2018-01,FTAG-2020-08,ATLAS-CONF-2018-006}.
Subdominant uncertainties include uncertainties related to the soft term in
the \met\ calculation~\cite{PERF-2016-07} and to the \met energy scale and resolution,
uncertainties in the reweighting of the MC event samples to match the pile-up conditions in data,
and a 1.7\%~\cite{ATLAS-CONF-2019-021}
uncertainty in the integrated luminosity of the combined 2015--2018 dataset.

Theoretical uncertainties include cross-section and other modelling uncertainties for all
background samples. The cross-section uncertainties considered are of  order  5\% to 6\% for the
\zjets, \ttbar, and \VV samples~\cite{TOPQ-2018-08,HIGG-2018-28} and of order 10\%~\cite{TOPQ-2018-08} \ttbarZ samples. A conservative
50\% uncertainty is used for the \ttbarW sample.
Uncertainties due to the choice of generator
or showering algorithm are estimated using additional
samples from alternative generators.
The uncertainties due to the choice of factorization and renormalization scales,
the modelling of ISR and FSR, and the choice of PDF set are estimated by
varying the nominal sample via additional generator weights, if available, or by using alternative samples.
An additional shape uncertainty is applied to the modelling of the
heavy-flavour fraction (HF) of jets in the \zjets and \VV samples. This uncertainty is
estimated by separating, using generator-level information,
the events with jets originating from $b$ or $c$ quarks from events
with jets originating from light quarks and rescaling the HF contribution by 50\%.
In the \ppdil channel, the HF uncertainty is uncorrelated between the categories
requiring exactly one or at least two \btagged jets.  In the \pptril channel,
the HF uncertainty is applied in a correlated way across the SR categories and the \VV CR,  since both regions
have the same $b$-tagging requirements.
Uncertainties in the VLQ pair-production cross-section similarly include the independent variation of the
factorization and renormalization scales and variations in the PDF and strong coupling constant,
following the PDF4LHC~\cite{Butterworth:2015oua} prescription with the MSTW 2008 68\% CL NNLO,
CT10 NNLO~\cite{Lai:2010vv,Gao:2013xoa} and NNPDF2.3 5f FFN PDF sets.

\section{Results}

The compatibility of the data with the background-only hypothesis
or with the signal-plus-background hypothesis
is tested
with a binned likelihood fit of
the discriminating variables in the categories of each individual channel.
The uncertainties are included as nuisance parameters (NP) with Gaussian
constraints in the likelihood fit.  Additional NPs are included
to take into account the statistical uncertainties in each bin
for each event category due to the limited size of the simulated samples.
The likelihood function $L(\mu,\vec{\theta})$
is constructed as a product of Poisson probabilities for each bin in
the discriminating variable in each category and depends on the signal-strength
factor $\mu$, which multiplies the expected signal cross-section $\sigma$,
and the set of all NPs $\vec{\theta}$.
Test statistics are based on the profile-likelihood ratio
$\lambda_\mu =  L(\mu,\hat{\vec{\theta}}_\mu)/L(\hat{\mu},\hat{\vec{\theta}}_{\hat{\mu}})$
where $\hat{\mu}$ and $\hat{\vec{\theta}}_{\hat{\mu}}$ are the values of $\mu$ and $\vec{\theta}$
that maximize the likelihood function, and $\hat{\vec{\theta}}_{\mu}$ are the
values of $\vec{\theta}$ that maximize the likelihood for a given $\mu$~\cite{Cowan:2010js}. Hypothesis tests
are performed with \textsc{RooStats}~\cite{Moneta:2010pm} with statistical models implemented using \textsc{RooFit}~\cite{RooFit} and \textsc{HistFactory}~\cite{Cranmer:1456844}.
The fit procedure was first validated in pseudo-data obtained by the sum of
the expected background contributions. Subsequently, the procedure
and the modelling of the background were tested by
including in the fit those categories in which the benchmark
signal contribution is expected to be smaller than 3\% of
the total background. Before the final full unblinding of the data,
an additional check was performed by including in the fit
all bins in all analysis categories with expected signal
contributions smaller than 5\% of the total background in those bins.

\begin{figure}[htbp]
\centering
\includegraphics[width=\linewidth]{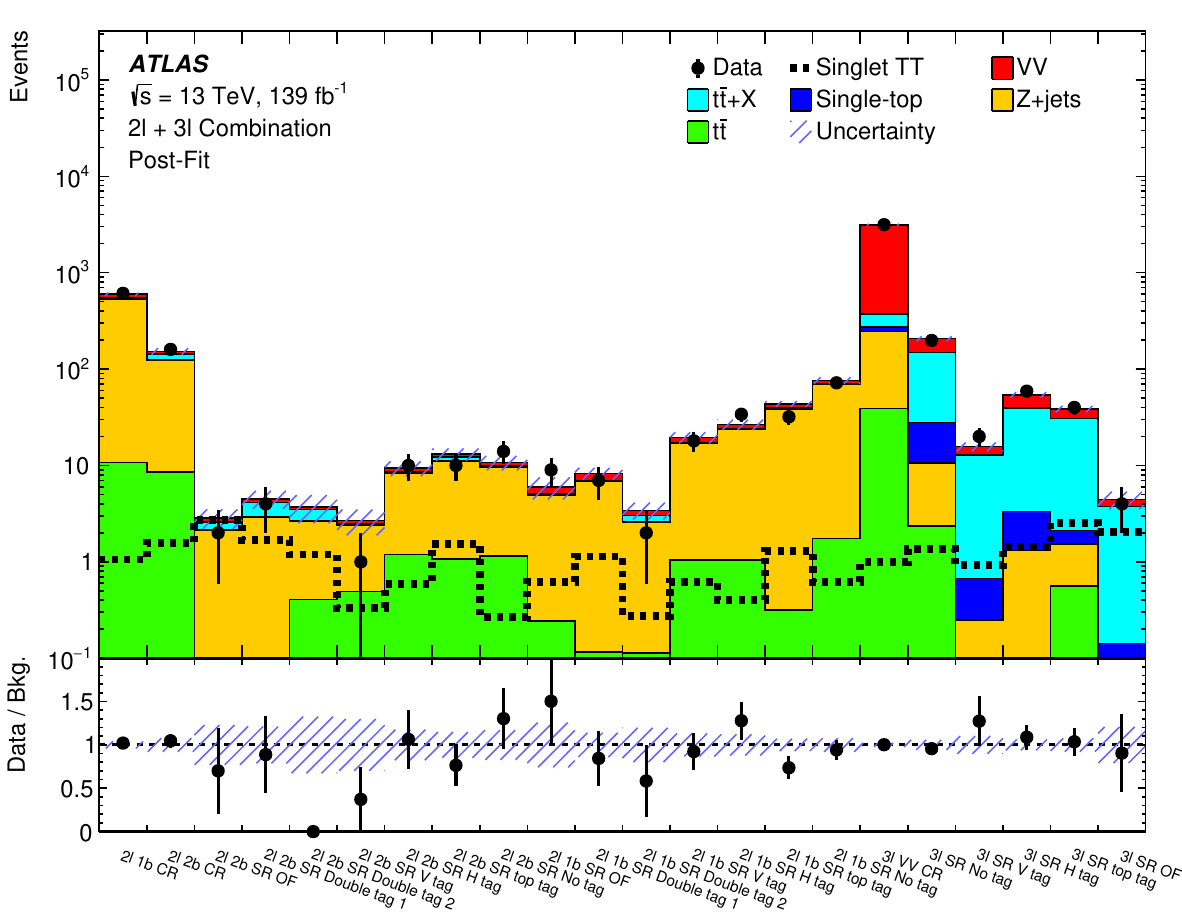}
\caption{Summary of data and background yields in all analysis categories
after the background-only fit.
The expected yields for singlet \TTbar signal with $m_T=\SI{1.2}{\TeV}$ are also shown for comparison.
The bottom panel shows the ratio of the data over the total background prediction.
The shaded band includes statistical and post-fit systematic uncertainties.}
\label{fig:summaryPlots}
\end{figure}

\begin{figure}[!htbp]
\centering
\subfloat[]{
\includegraphics[width=.3\linewidth]{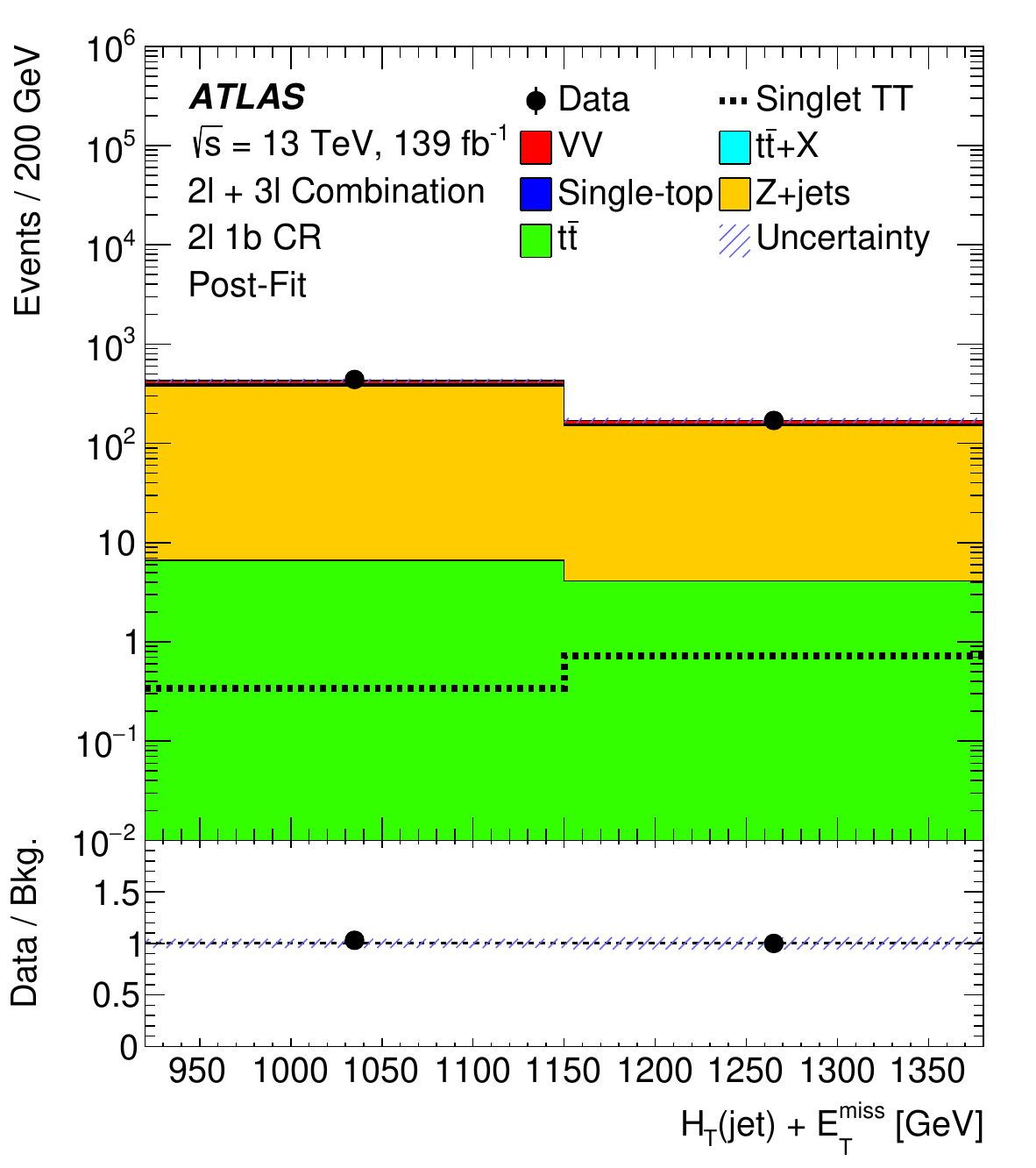}
}
\subfloat[]{
\includegraphics[width=.3\linewidth]{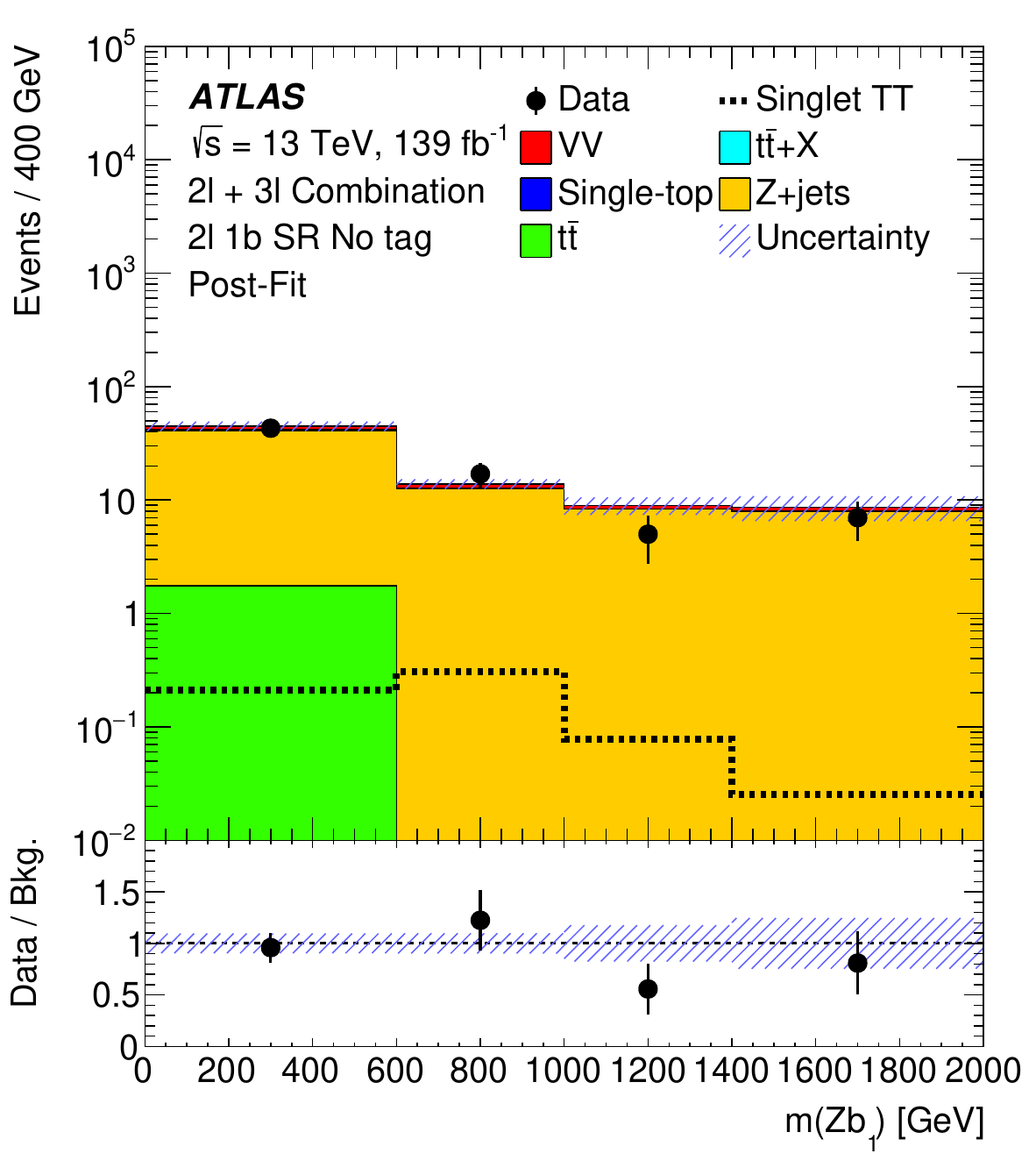}
}
\subfloat[]{
\includegraphics[width=.3\linewidth]{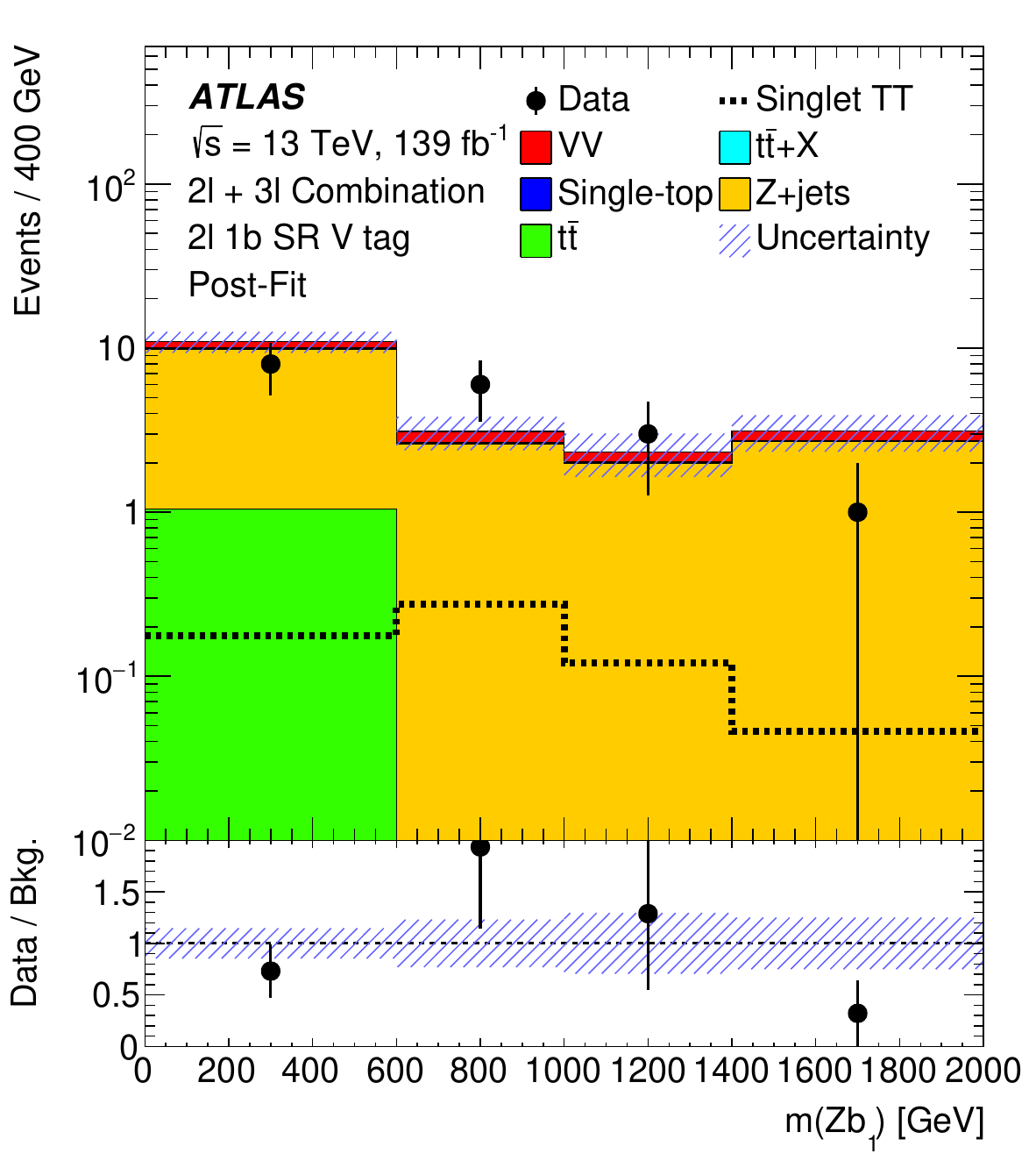}
}\\
\subfloat[]{
\includegraphics[width=.3\linewidth]{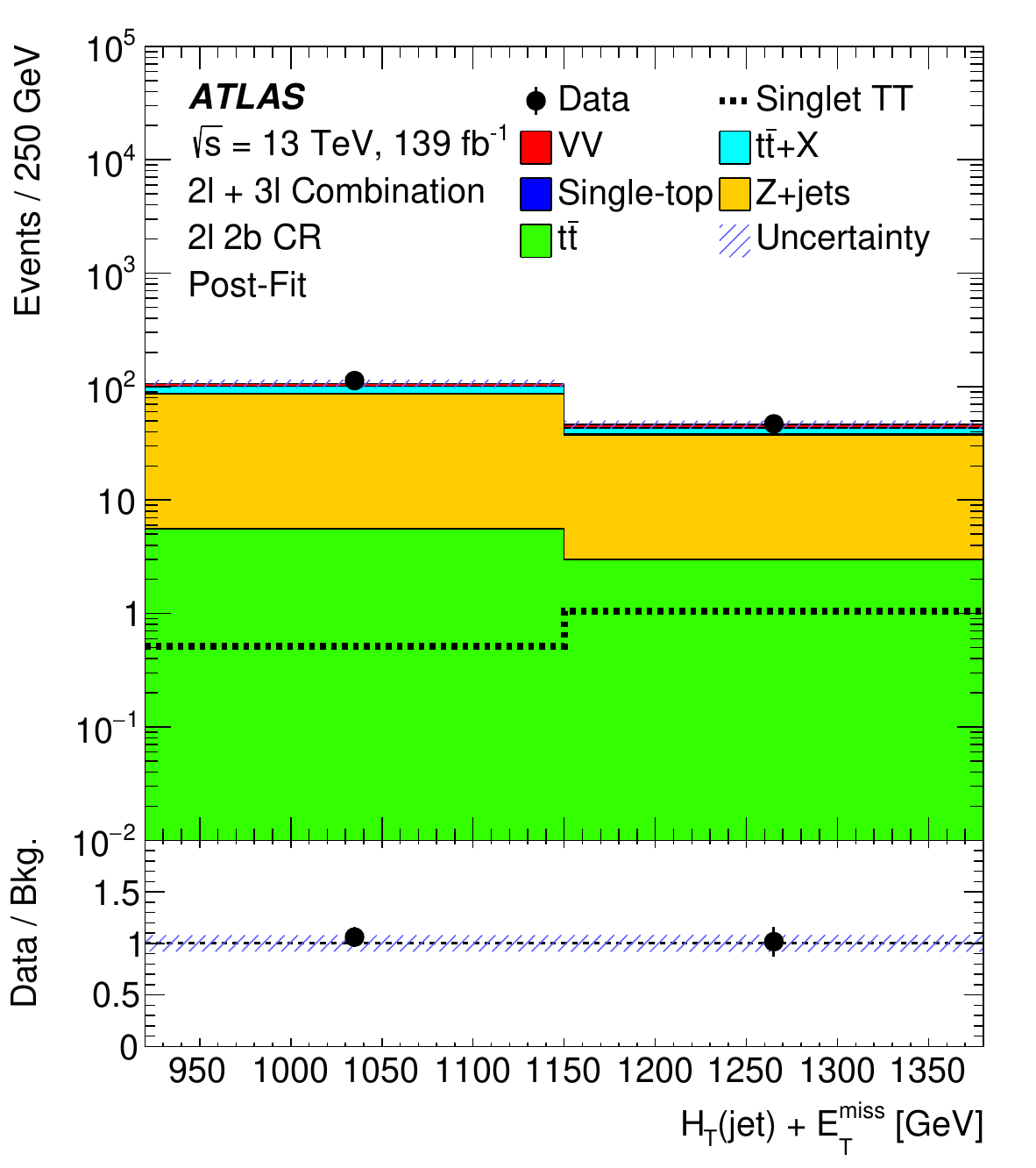}
}
\subfloat[]{
\includegraphics[width=.3\linewidth]{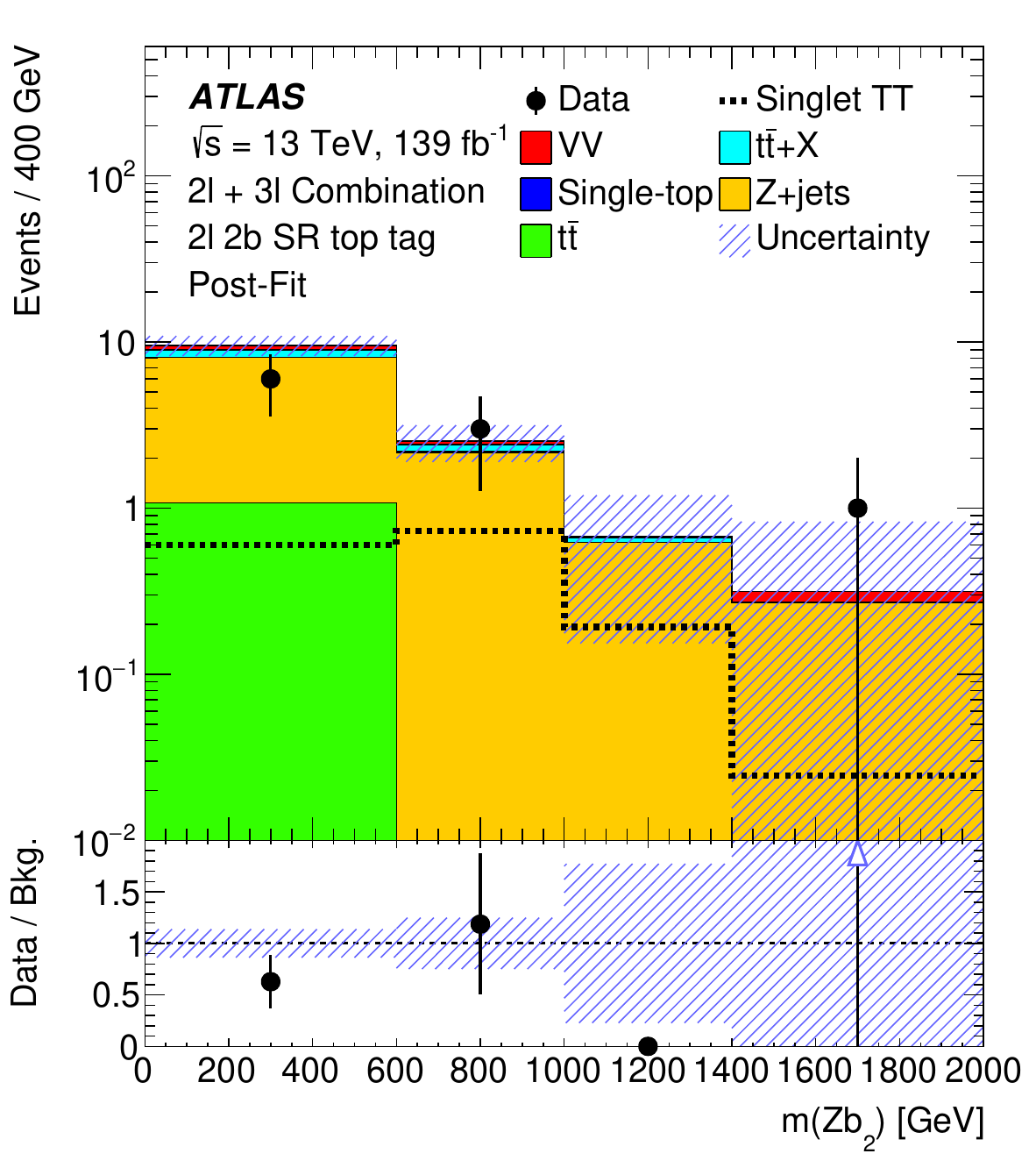}
}
\subfloat[]{
\includegraphics[width=.3\linewidth]{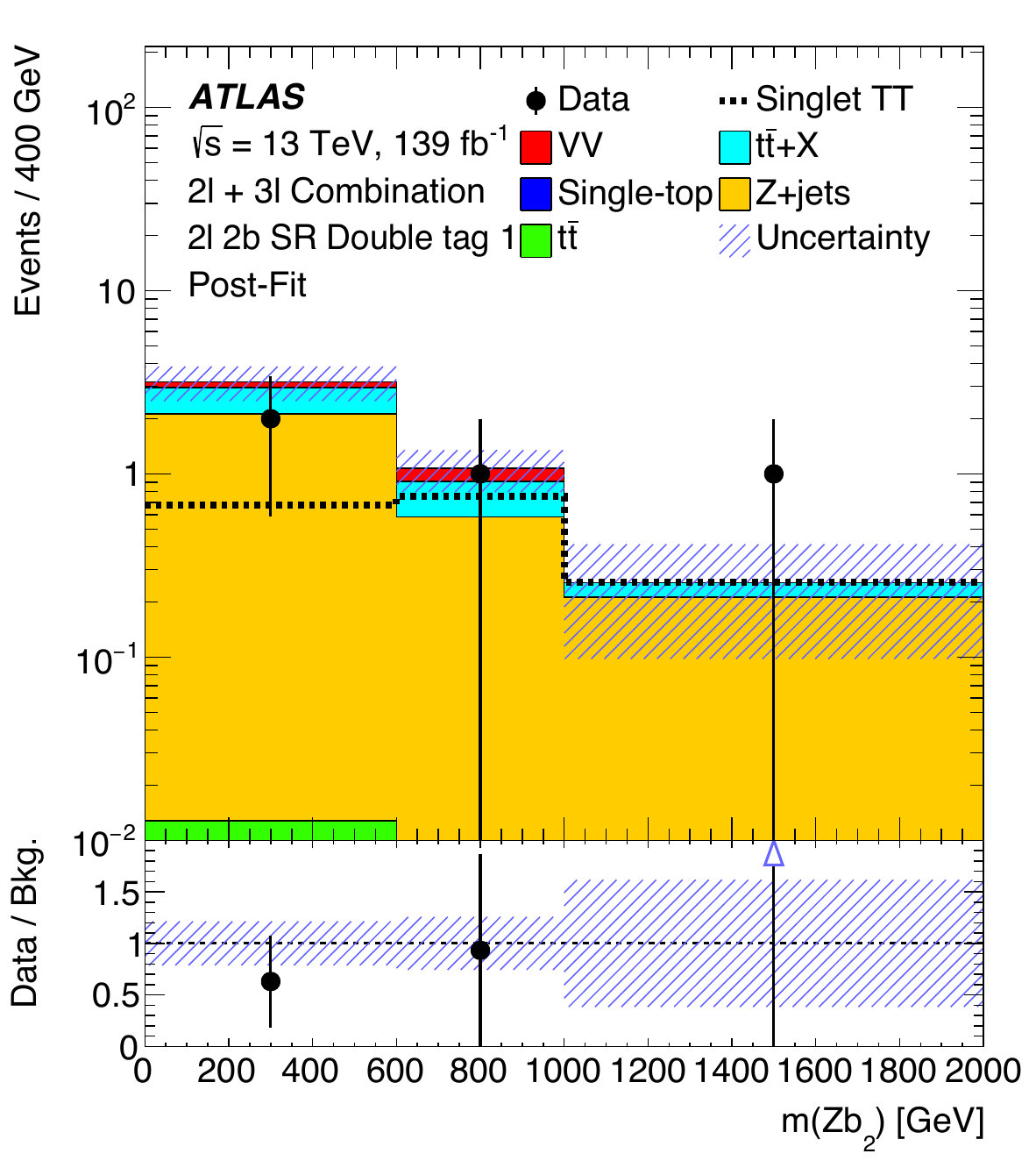}
}
 
\caption{Distributions of the final discriminants for the \ppdil channel
in the control regions and in selected event categories of the signal regions:
(a) control region and (b,c) categories requiring exactly one $b$-jet or
(d) control region and (e,f) categories requiring at least two $b$-jets.
The distributions are shown after the combined background-only fit.
The distributions expected for singlet \TTbar signal with $m_T=\SI{1.2}{\TeV}$ are
also shown in overlay. The bottom panels show the ratio of the data over the total background prediction.
The two blue triangles indicate points beyond the vertical range. The last bin contains the overflow.}
\label{fig:exampleRegionsDil}
\end{figure}

\begin{figure}[!htbp]
\centering
\subfloat[]{
\includegraphics[width=.3\linewidth]{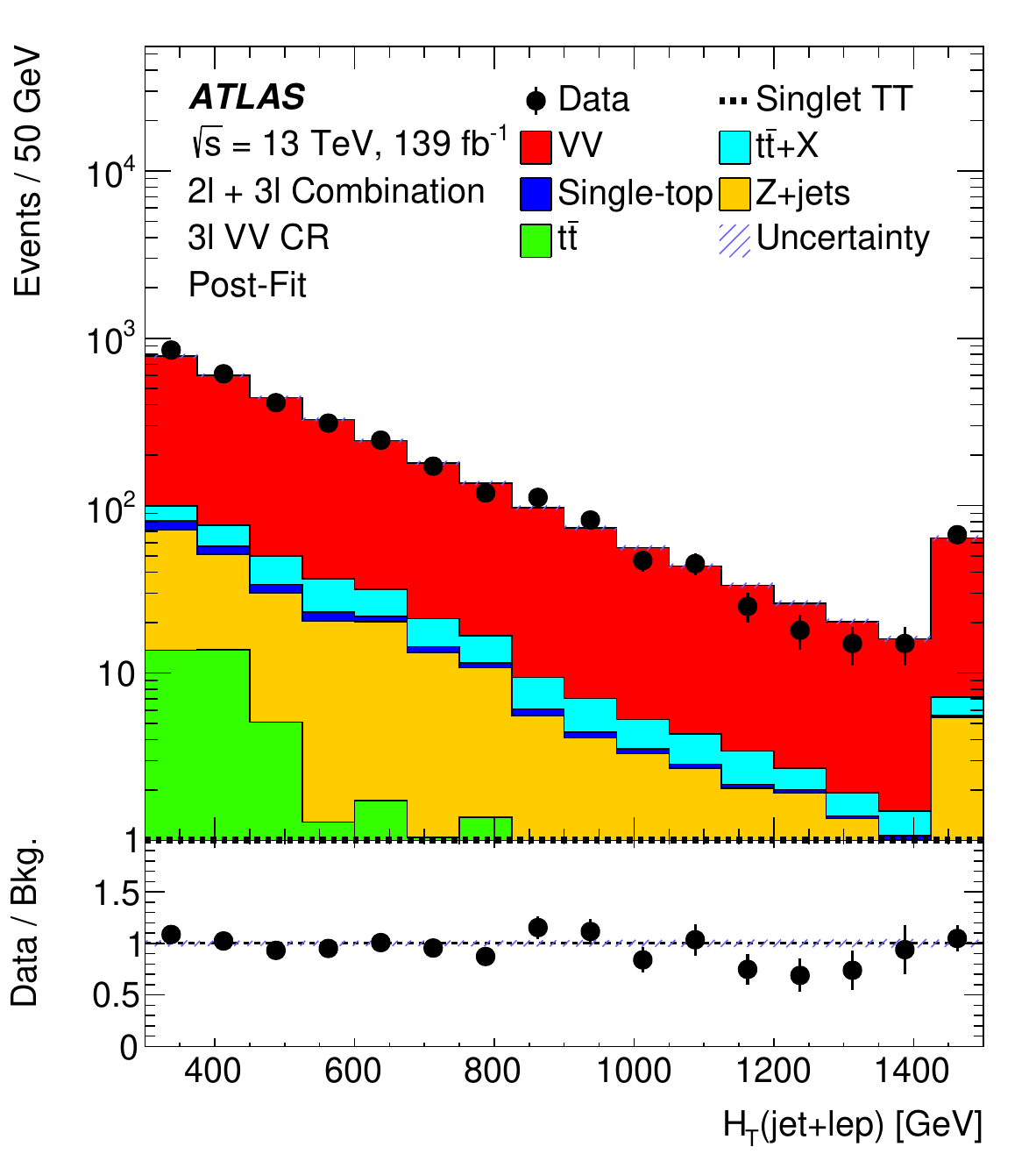}
}
\subfloat[]{
\includegraphics[width=.3\linewidth]{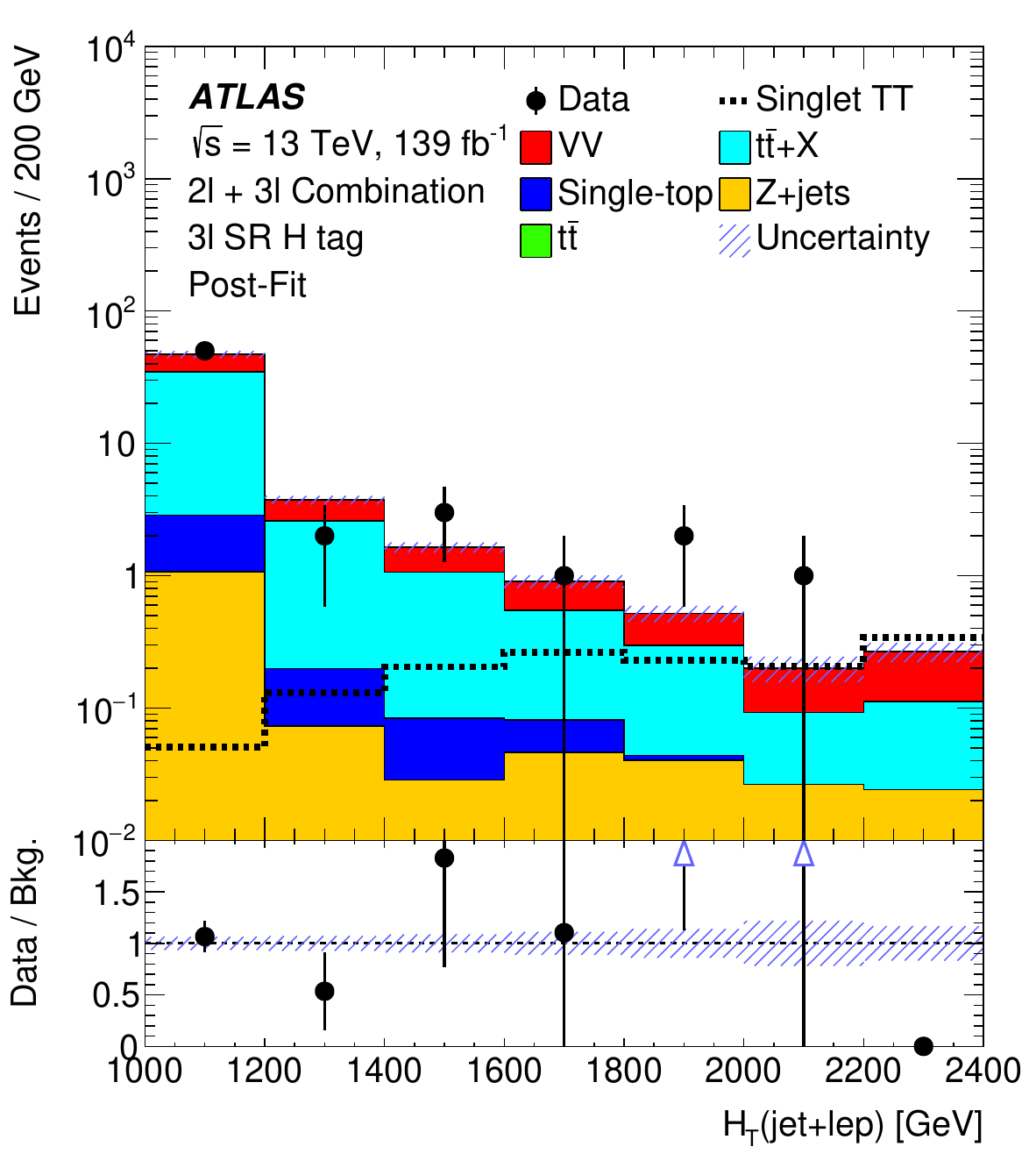}
}
\subfloat[]{
\includegraphics[width=.3\linewidth]{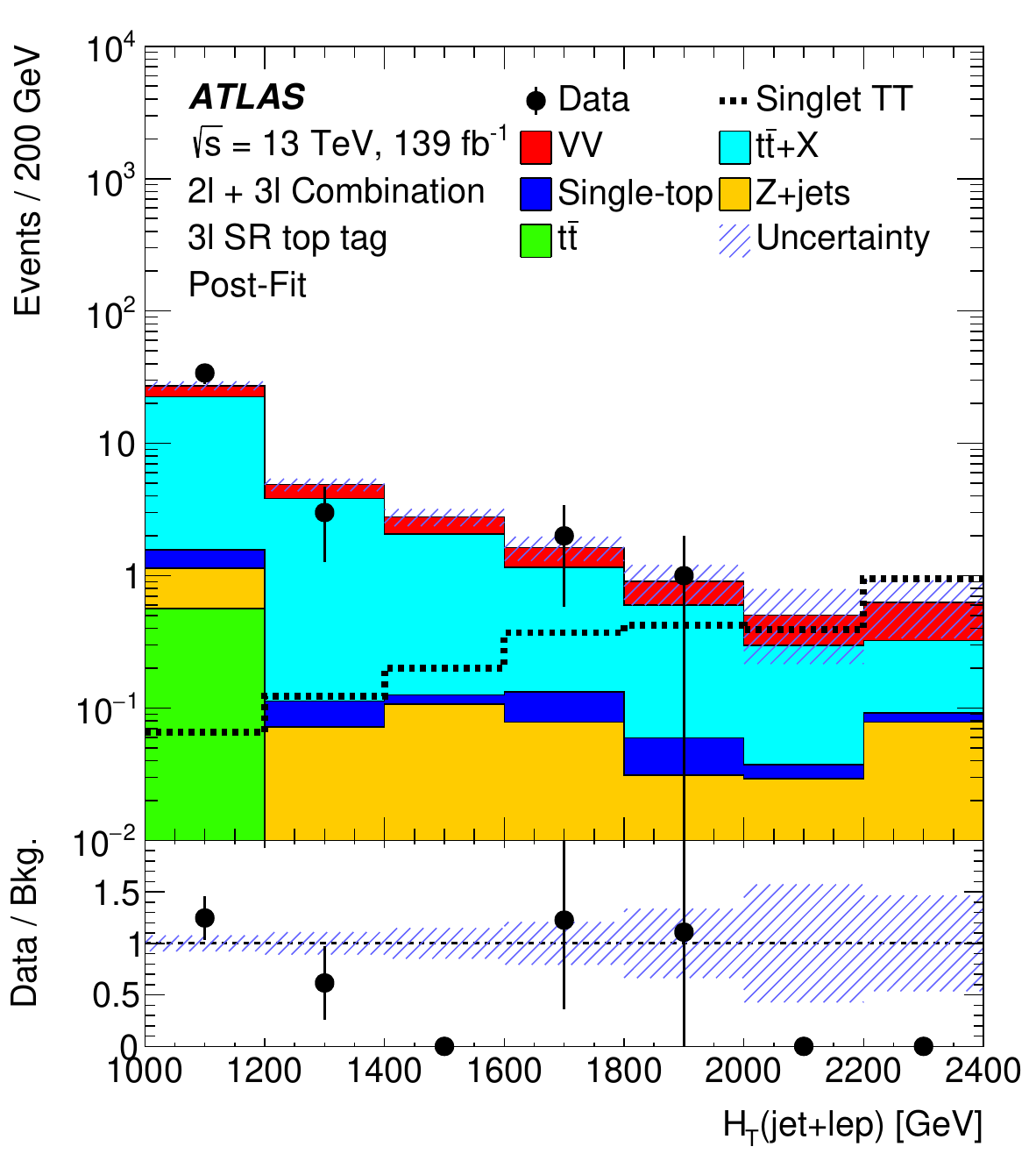}
}

\caption{Distributions of the final discriminants for the \pptril channel
in the (a) $VV$ control region and in (b,c) selected
event categories of the signal region.
The distributions are shown after the combined background-only fit.
The distributions expected for singlet \TTbar signal with $m_T=\SI{1.2}{\TeV}$ are
also shown in overlay. The bottom panels show the ratio of the data over the total background prediction.
The two blue triangles indicate points beyond the vertical range. The last bin contains the overflow.}
\label{fig:exampleRegionsTril}
\end{figure}

A background-only fit to the data, in which $\mu$ is set to zero,
is performed using $q_0$~\cite{Cowan:2010js} as the test statistic and no
significant excess over the background expectation is observed.
To obtain the final result,  a simultaneous fit of all
the regions and categories of both channels is performed.
The results are found to be insensitive to variations
of the correlation model and the following approach was adopted:
all experimental
and cross-section uncertainties are treated as fully correlated between the two analysis
channels, while the remaining, modelling-related, uncertainties are treated as uncorrelated.
\Fig{\ref{fig:summaryPlots}} shows
the background and observed data yields in
all the analysis categories after the
combined background-only fit (post-fit).  Post-fit distributions of the fitted variables
in the background control regions and in selected
signal region categories are shown in \Fig{\ref{fig:exampleRegionsDil}} for the
\ppdil channel and in \Fig{\ref{fig:exampleRegionsTril}} for the \pptril channel.
The figures include the expected distributions for \TTbar signal with $m_T=\SI{1.2}{\TeV}$
generated in the singlet configuration. The event yields for the dominant backgrounds
in each event category after the background-only fit are also summarized in Table~\ref{tab:yields_postfit}, which
includes the number of events observed in data. The table also includes the expected yields in number of events
for \TTbar and \BBbar production in the singlet configuration with $m_{\mathrm{VLQ}}=\SI{1.2}{\TeV}$.
These yields and their uncertainties are provided before any fit is performed (pre-fit).

\begin{table}[htbp]
\begin{center}
\caption{Summary of observed and predicted yields in the control regions
and signal region categories. The background prediction is shown after the
combined likelihood fit to data under the background-only hypothesis
across all control region and signal region categories.
The uncertainties include statistical and systematic uncertainties.
Due to correlations, the total background uncertainty is not necessarily
equal to the sum in quadrature of the uncertainties of the individual background processes.
The expected yields for benchmark signals with $m_{\mathrm{VLQ}}=\SI{1.2}{\TeV}$
obtained by using their theoretical cross-sections
are also shown with their pre-fit uncertainties. Dashes indicate negligible yields.
}
\label{tab:yields_postfit}
\resizebox{\textwidth}{!}{
\begin{tabular}{ l|
S[table-format=4, separate-uncertainty]      |      
S[table-format=4.1(3), separate-uncertainty] ||       
S[table-format=4.2(2), separate-uncertainty]        
S[table-format=3.2(2), separate-uncertainty]        
S[table-format=3.3(3), separate-uncertainty]     
S[table-format=3.2(2), separate-uncertainty]        
S[table-format=3.2(3), separate-uncertainty] ||       
S[table-format=1.2(3), separate-uncertainty]        
S[table-format=1.2(3), separate-uncertainty]}        
\toprule{{Event category}} & {{Data}}    &{{Total background}}   & {{$VV$}}    & {{$t\bar{t}+X$}}& {{Single-top}} & {{$Z$+jets}}    & {{$t\bar{t}$}}  & {{Singlet \TTbar}}  & {{Singlet \BBbar}}  \\
& & & & & & & & {{$m_T=\SI{1.2}{\TeV}$}} & {{$m_B=\SI{1.2}{\TeV}$}} \\ \midrule\midrule{$2\ell$ $1b$ CR}              & 610   & 598 \pm 24       & 50 \pm 18     & 13 \pm 4      & 1.3 \pm 0.6     & 522 \pm 30    & 11 \pm 4     & 1.06 \pm 0.08 & 1.27 \pm 0.09   \\
{$2\ell$ $1b$ SR No tag}       & 72    & 76 \pm 6         & 5.5 \pm 2.0   & 0.37 \pm 0.13 & 0.8 \pm 0.9     & 68 \pm 6      & 1.8 \pm 1.9  & 0.62 \pm 0.13 & 1.52 \pm 0.23   \\
{$2\ell$ $1b$ SR V tag}        & 18    & 19.6 \pm 2.6     & 2.2 \pm 0.8   & 0.22 \pm 0.08 & {--}            & 16.1 \pm 2.1  & 1.1 \pm 1.3  & 0.63 \pm 0.07 & 1.31 \pm 0.13   \\
{$2\ell$ $1b$ SR H tag}        & 34    & 26.6 \pm 3.3     & 2.4 \pm 0.9   & 0.38 \pm 0.09 & {--}            & 22.8 \pm 3.0  & 1.1 \pm 1.5  & 0.41 \pm 0.05 & 0.66 \pm 0.09   \\
{$2\ell$ $1b$ SR top tag}      & 32    & 43.5 \pm 3.1     & 3.6 \pm 1.3   & 1.25 \pm 0.22 & {--}            & 38.3 \pm 3.3  & 0.3 \pm 0.6  & 1.29 \pm 0.14 & 1.40 \pm 0.14   \\
{$2\ell$ $1b$ SR Double tag 1} & 7     & 8.3 \pm 1.2      & 1.3 \pm 0.5   & 0.20 \pm 0.09 & {--}            & 6.7 \pm 1.0   & 0.1 \pm 0.4  & 1.14 \pm 0.15 & 1.05 \pm 0.17   \\
{$2\ell$ $1b$ SR Double tag 2} & 2     & 3.4 \pm 0.7      & 0.39 \pm 0.19 & 0.42 \pm 0.10 & {--}            & 2.5 \pm 0.5   & 0.11 \pm 0.31& 0.27 \pm 0.06 & 0.19 \pm 0.06   \\
{$2\ell$ $1b$ SR OF}     & 9     & 6.0 \pm 1.6      & 0.9 \pm 0.4   & 0.18 \pm 0.05 & {--}            & 4.7 \pm 1.3   & 0.2 \pm 0.7  & 0.62 \pm 0.08 & 0.60 \pm 0.11   \\
\midrule{$2\ell$ $2b$ CR}              & 160   & 152 \pm 12       & 9 \pm 4       & 18 \pm 6      & 1.2 \pm 0.6     & 115 \pm 13    & 9.0 \pm 3.5  & 1.58 \pm 0.10 & 1.90 \pm 0.13   \\
{$2\ell$ $2b$ SR No tag}       & 14    & 10.7 \pm 1.9     & 0.9 \pm 0.4   & 0.07 \pm 0.08 & 0.3 \pm 0.5     & 8.3 \pm 1.1   & 1.2 \pm 1.4  & 0.27 \pm 0.07 & 1.11 \pm 0.17   \\
{$2\ell$ $2b$ SR V tag}        & 1     & 2.7 \pm 0.8      & 0.22 \pm 0.13 & 0.06 \pm 0.05 & {--}            & 1.9 \pm 0.5   & 0.5 \pm 0.7  & 0.34 \pm 0.04 & 0.81 \pm 0.08   \\
{$2\ell$ $2b$ SR H tag}        & 10    & 9.4 \pm 1.6      & 0.76 \pm 0.34 & 0.26 \pm 0.09 & {--}            & 7.1 \pm 1.0   & 1.2 \pm 1.4  & 0.59 \pm 0.09 & 1.90 \pm 0.16   \\
{$2\ell$ $2b$ SR top tag}      & 10    & 13.0 \pm 1.9     & 0.8 \pm 0.4   & 1.10 \pm 0.30 & {--}            & 10.0 \pm 1.6  & 1.1 \pm 1.2  & 1.56 \pm 0.18 & 1.97 \pm 0.15   \\
{$2\ell$ $2b$ SR Double tag 1} & 4     & 4.5 \pm 1.0      & 0.40 \pm 0.25 & 1.2 \pm 0.6   & {--}            & 2.9 \pm 0.8   & 0.01 \pm 0.19& 1.71 \pm 0.11 & 1.71 \pm 0.18   \\
{$2\ell$ $2b$ SR Double tag 2} & 0     & 3.7 \pm 1.2      & 0.25 \pm 0.14 & 0.78 \pm 0.18 & {--}            & 2.3 \pm 0.6   & 0.4 \pm 1.1  & 1.22 \pm 0.10 & 1.12 \pm 0.10   \\
{$2\ell$ $2b$ SR OF}     & 2     & 2.8 \pm 0.7      & 0.28 \pm 0.15 & 0.44 \pm 0.20 & {--}            & 2.1 \pm 0.6   & {--}         & 2.76 \pm 0.25 & 2.02 \pm 0.26   \\
\midrule{$3\ell$ VV CR}              & 3149  & 3140 \pm 70      & 2770 \pm 90   & 98 \pm 9      & 26.7 \pm 1.6    & 210 \pm 60    & 40 \pm 18    & 1.00 \pm 0.14 & 0.90 \pm 0.14   \\
{$3\ell$ SR No tag}          & 198   & 203 \pm 11       & 59 \pm 5      & 117 \pm 10    & 17.3 \pm 0.6    & 8.5 \pm 3.0   & 2.4 \pm 2.0  & 1.37 \pm 0.17 & 1.31 \pm 0.17   \\
{$3\ell$ SR V tag}           & 20    & 15.3 \pm 1.5     & 2.82 \pm 0.34 & 11.8 \pm 1.5  & 0.42 \pm 0.06   & 0.25 \pm 0.11 & {--}         & 0.94 \pm 0.07 & 0.71 \pm 0.09   \\
{$3\ell$ SR H tag}           & 59    & 52.9 \pm 2.8     & 14.9 \pm 1.2  & 34.7 \pm 2.7  & 2.03 \pm 0.14   & 1.3 \pm 0.4   & {--}         & 1.44 \pm 0.10 & 0.96 \pm 0.08   \\
{$3\ell$ SR top tag}         & 40    & 37.7 \pm 2.6     & 7.9 \pm 0.6   & 27.6 \pm 2.4  & 0.59 \pm 0.07   & 1.0 \pm 0.4   & 0.6 \pm 1.2  & 2.54 \pm 0.16 & 1.73 \pm 0.12   \\
{$3\ell$ SR OF}        & 4     & 4.3 \pm 0.9      & 0.68 \pm 0.11 & 3.4 \pm 0.8   & 0.07 \pm 0.03 & 0.07 \pm 0.05 & {--}         & 2.07 \pm 0.16 & 0.69 \pm 0.13   \\
\bottomrule
\end{tabular}
}
 
\end{center}
\end{table}

\begin{figure}[htbp]
\centering
\subfloat[]{
\includegraphics[width=.49\linewidth]{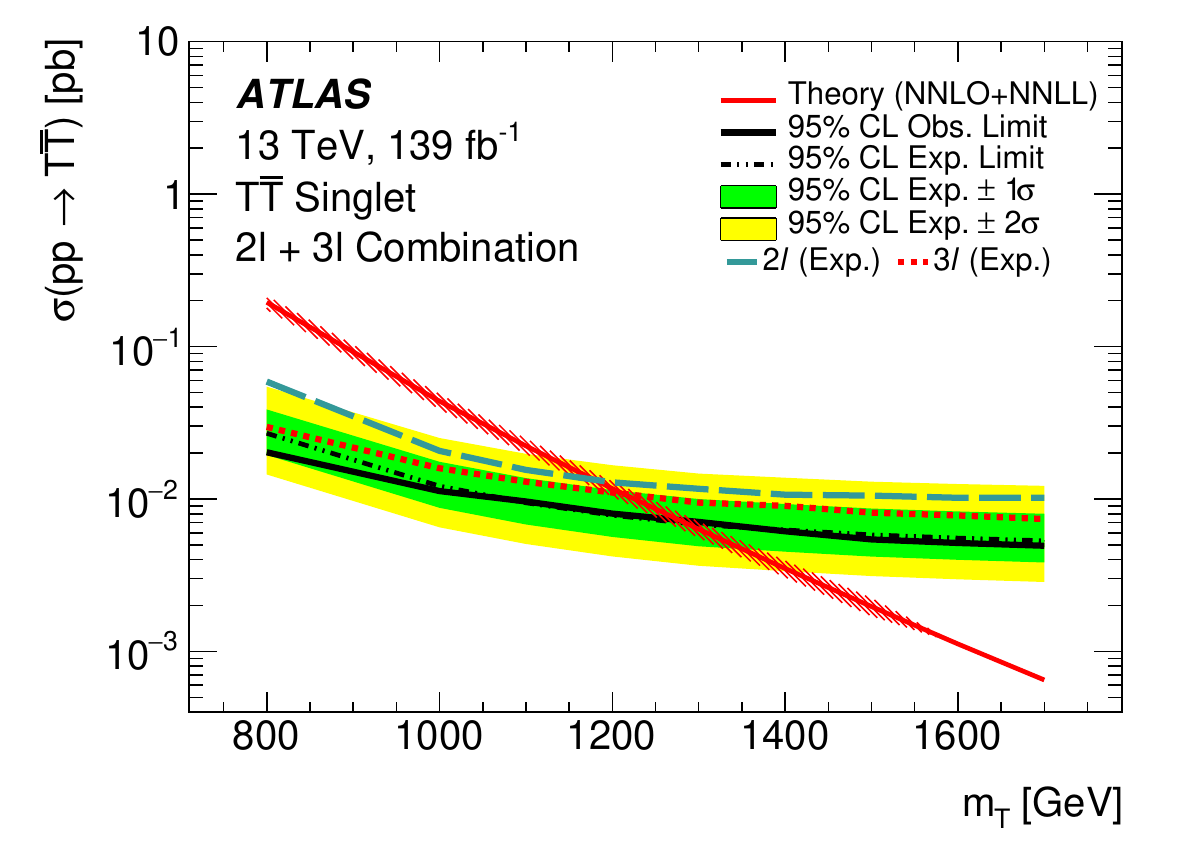}
}
\subfloat[]{
\includegraphics[width=.49\linewidth]{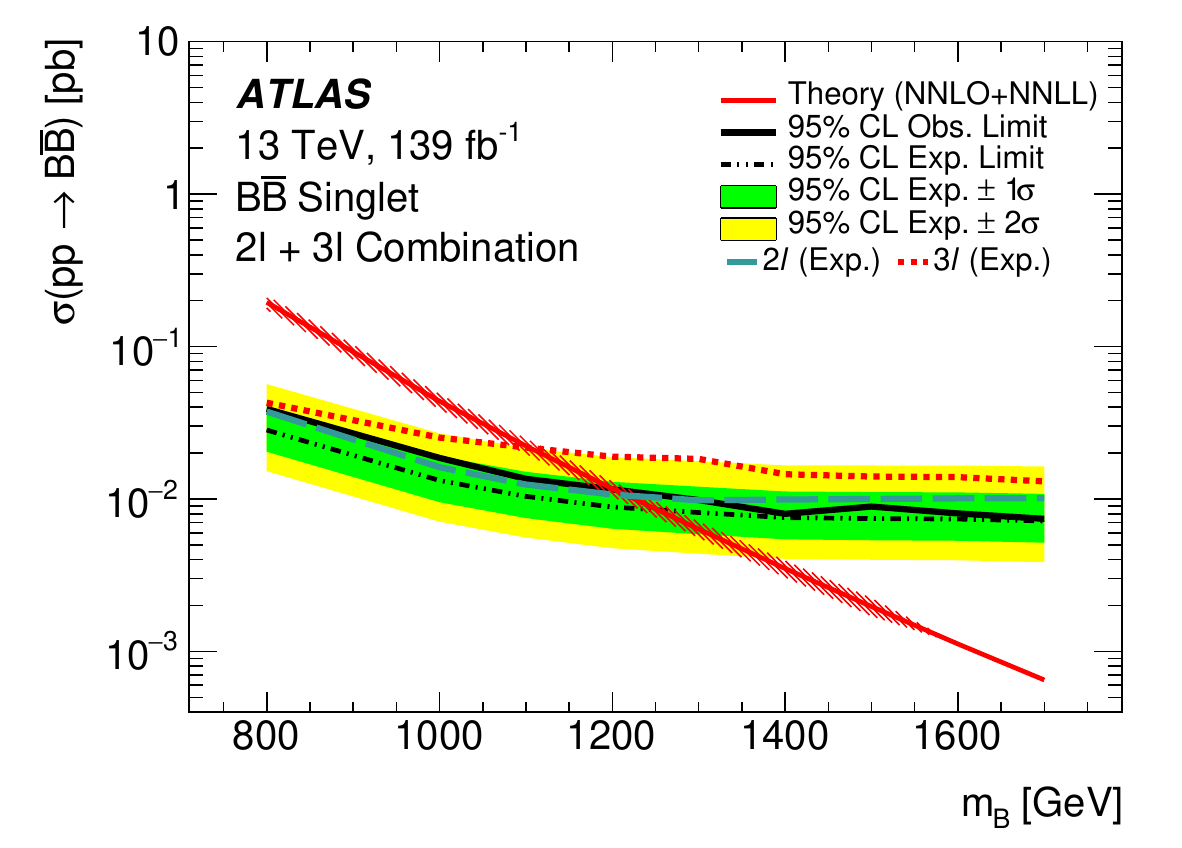}
}\\
\subfloat[]{
\includegraphics[width=.49\linewidth]{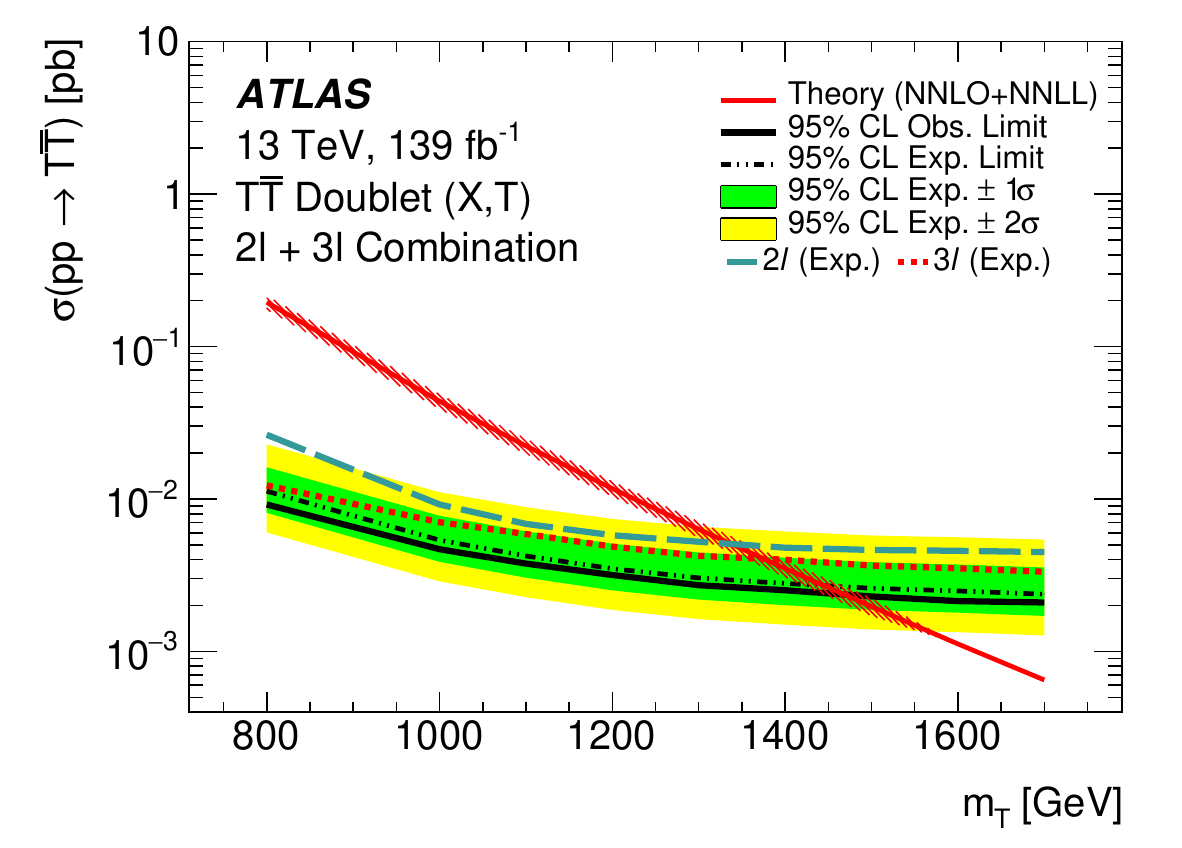}
}
\subfloat[]{
\includegraphics[width=.49\linewidth]{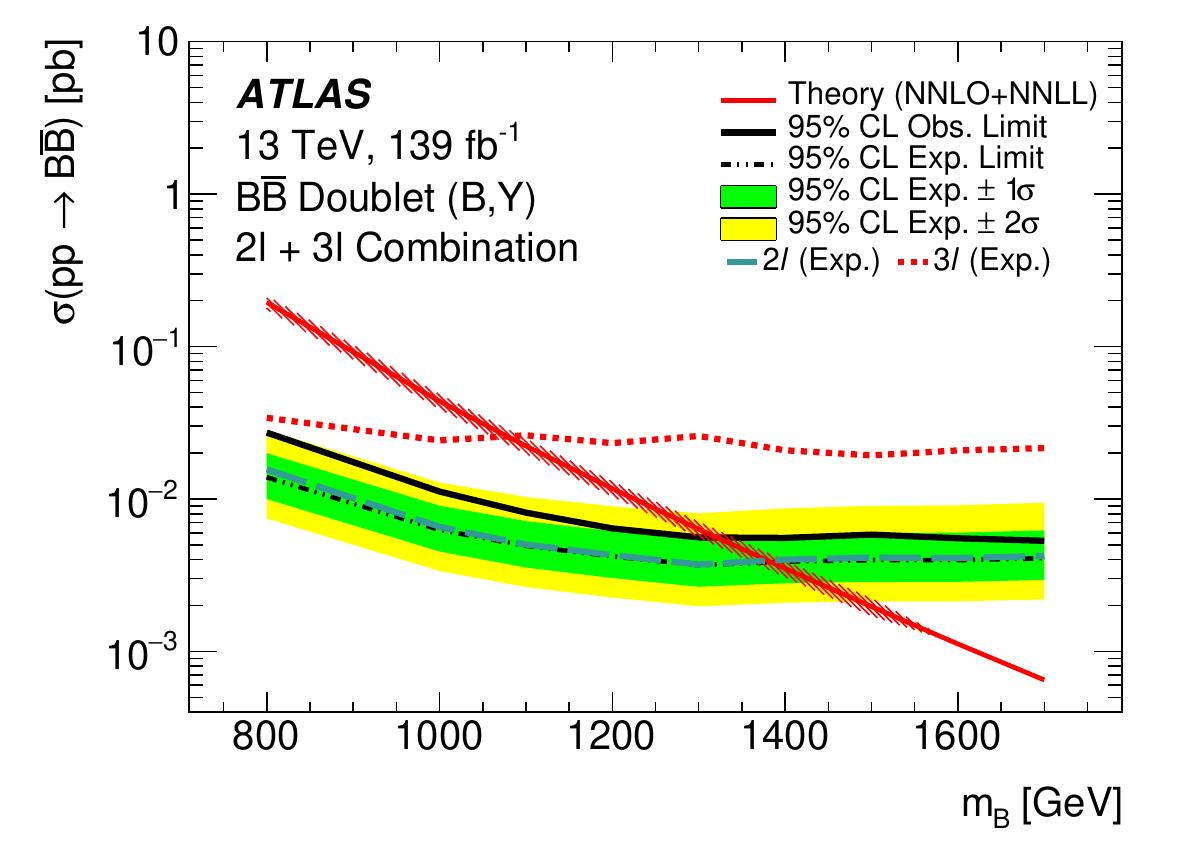}
}\\
\subfloat[]{
\includegraphics[width=.49\linewidth]{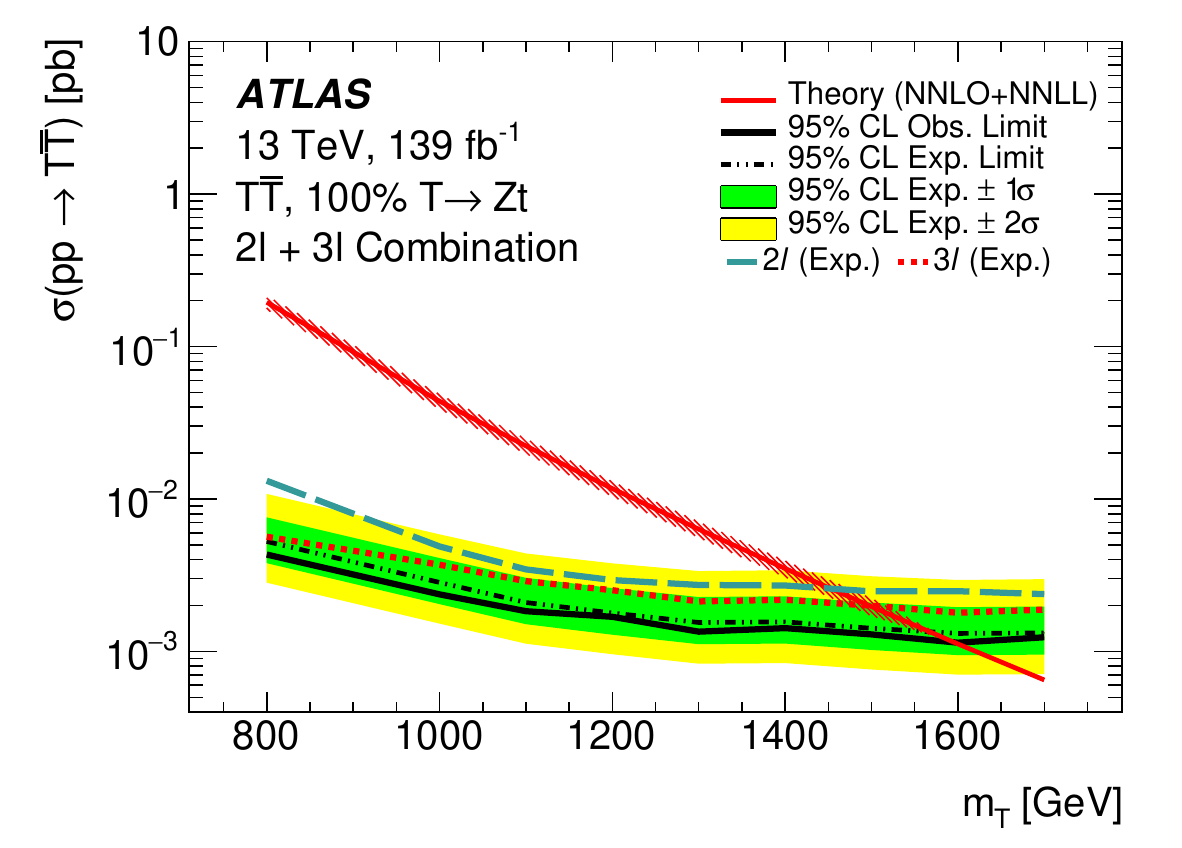}
}
\subfloat[]{
\includegraphics[width=.49\linewidth]{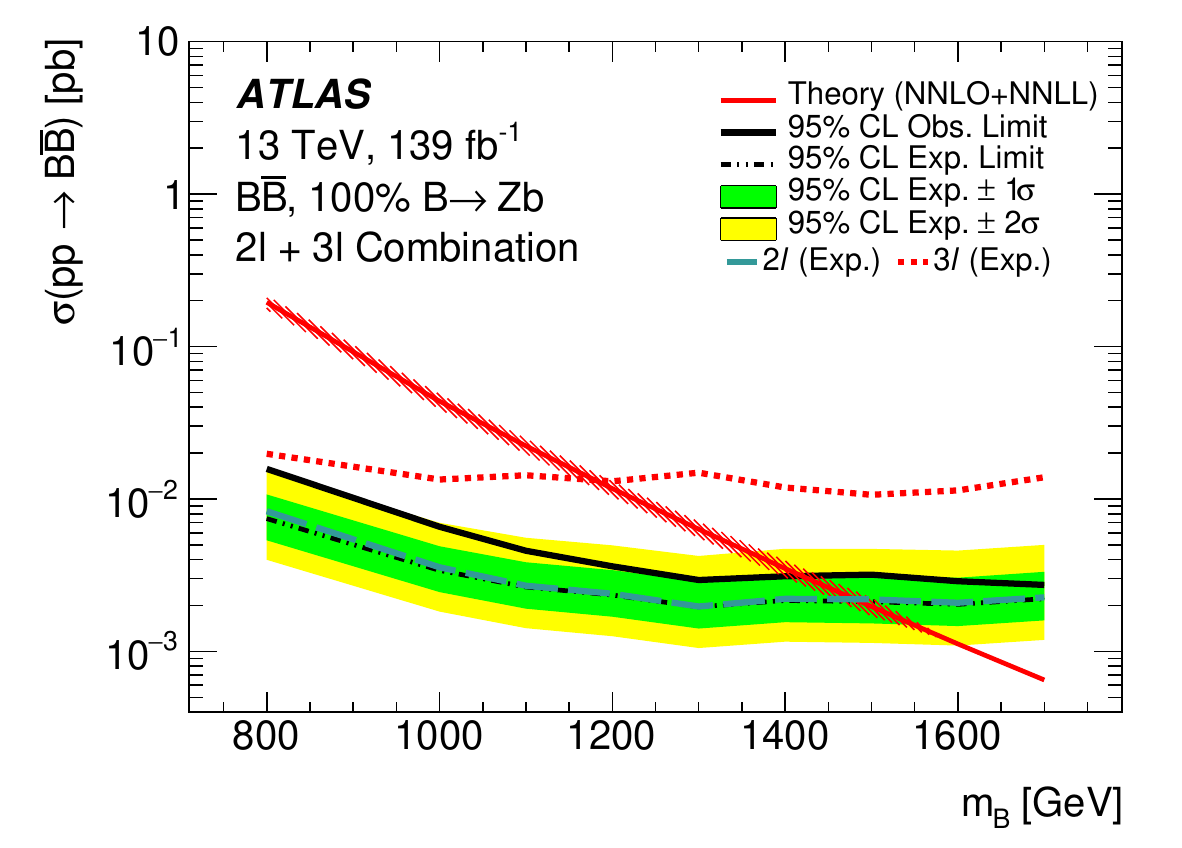}
}
\caption{Expected and observed combined limits at 95\% CL on the production cross-section for (a,c, and e) \TTbar and (b,d, and f) \BBbar  for three
representative \br compositions. The expected limits for the
two individual channels are also shown.}
\label{fig:ppcomb_limits}
\end{figure}

Upper limits on the pair-production cross-section as a function
of the VLQ mass are derived at 95\% CL for the $T$ and $B$ quarks.
These limits are derived by using $\tilde{q}_\mu$~\cite{Cowan:2010js} as the test statistic
with the CL$_\text{s}$ technique~\cite{Junk:1999kv,Read:2002hq}
in the asymptotic approximation~\cite{Cowan:2010js}.
The limits are found to agree within a few percent with the limits obtained with
pseudo-experiments.
For the $1\sigma$ uncertainty contours, the limits agree within 10\%, with
the $-1\sigma$ band being slightly too narrow in the asymptotic
approximation and the $+1\sigma$ band being slightly too wide.
The  expected and observed upper limits on the cross-section obtained from the combination
of the two channels are shown in \Fig{\ref{fig:ppcomb_limits}}.
To demonstrate the complementarity of the two channels, their
individually obtained expected limits are also shown.
Three benchmark scenarios for each of $B$ and $T$
are shown, namely the singlet, doublet, and $100\%$~\br configurations, with increasingly high \br to $Zb$ or $Zt$.
Comparisons with the respective theoretical predictions of the cross-sections
give the lowest-allowed VLQ masses as summarized in Table~\ref{tab:resultMassLimits}.
The higher signal purity in the \pptril channel results in an overall higher sensitivity for $T$
than in the \ppdil channel, while the larger data sample and finer categorization of the latter manifests
in a performance that is dominant in the results for $B$. For models with 100\% \br
to the $Z$ boson and a third-generation quark,
$T$ ($B$) masses up to $\SI{\tthundobs}{\TeV}$ ($\SI{\bbhundobs}{\TeV}$) are excluded at 95\% CL, while for the
singlet and doublet configurations, masses are excluded up to $\SI{\ttsobs}{\TeV}$ ($\SI{\bbsobs}{\TeV}$) and
up to $\SI{\ttsobs}{\TeV}$ ($\SI{\bbdobs}{\TeV}$), respectively. The combination of the two channels with
their fine categorization results in the observed sensitivity,
however, the limits are evidently driven by the categories
with high expected signal-to-background fractions,
namely, `2l 2b SR OF' (100\%) and `3l SR OF' (45\%).

\begin{figure}[!htbp]
\centering
\subfloat[]{
\includegraphics[width=.49\linewidth]{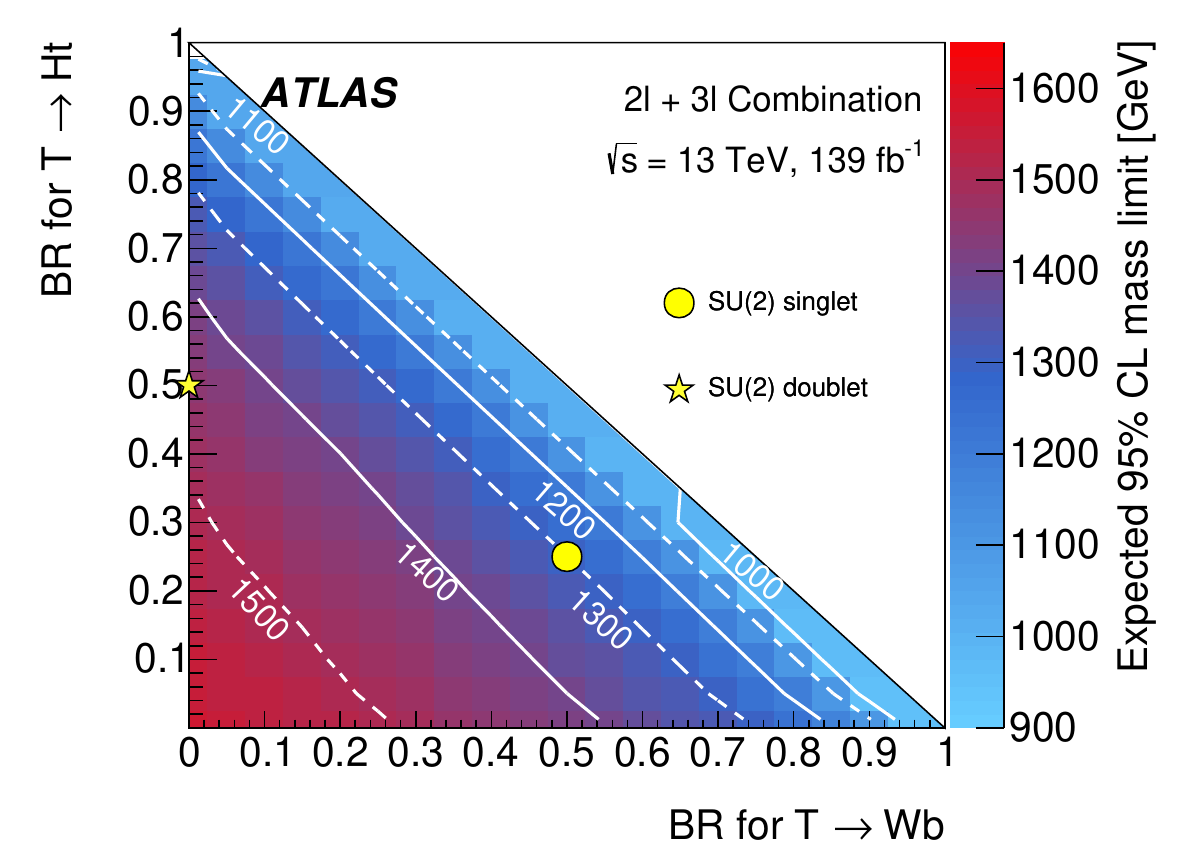}
}
\subfloat[]{
\includegraphics[width=.49\linewidth]{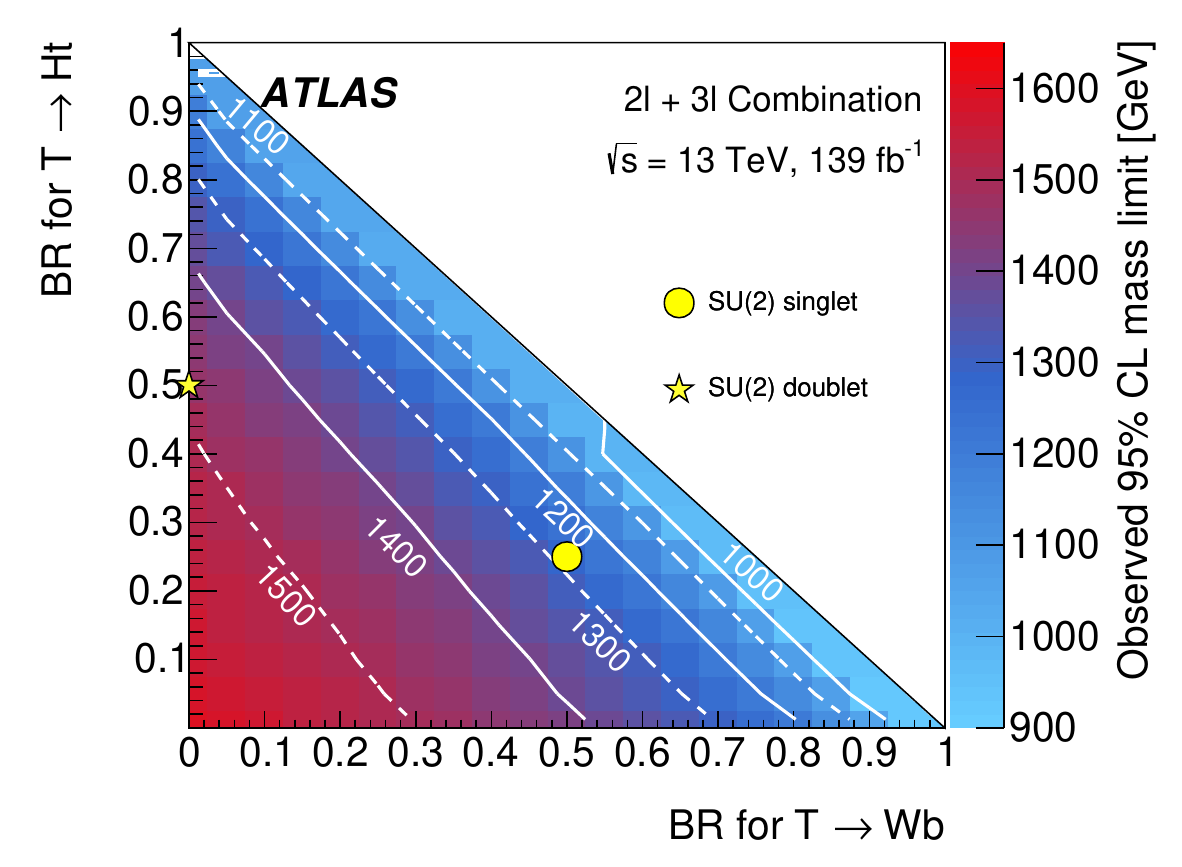}
 
}\\
\subfloat[]{
\includegraphics[width=.49\linewidth]{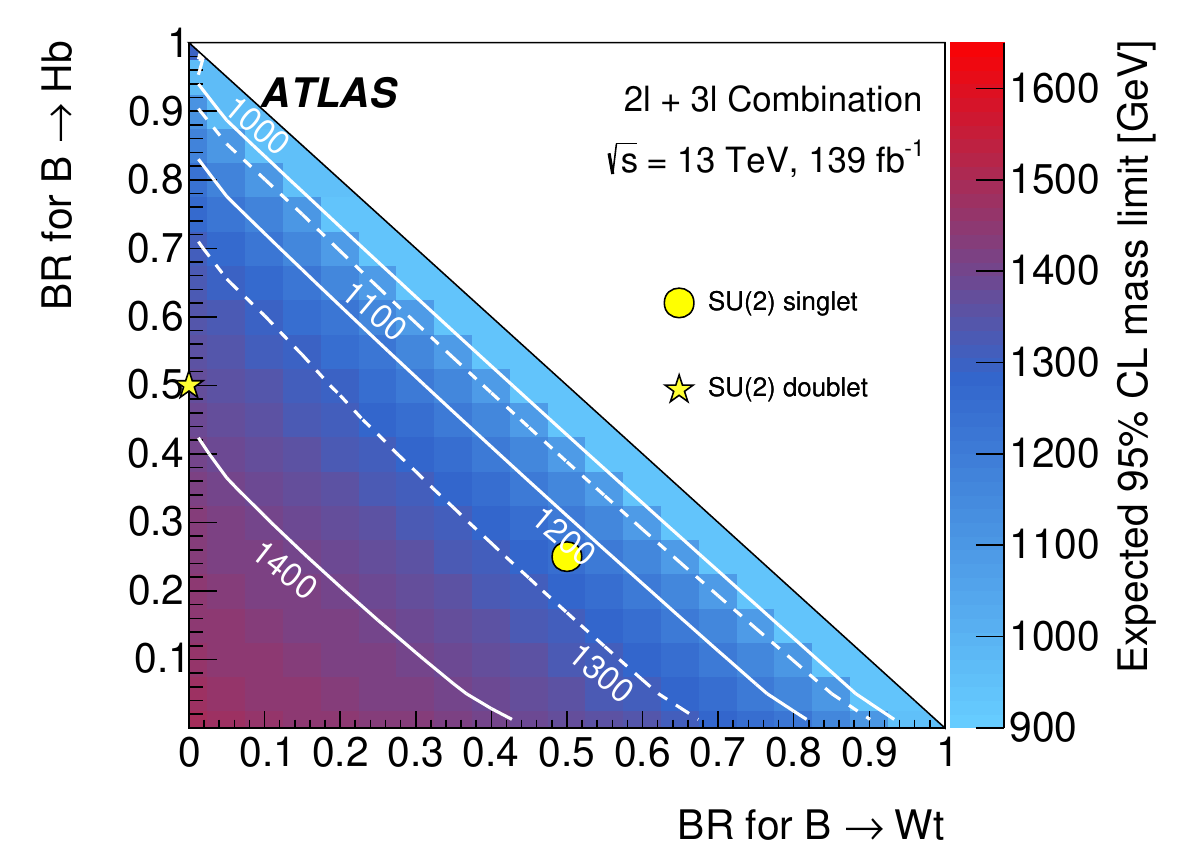}
 
}
\subfloat[]{
\includegraphics[width=.49\linewidth]{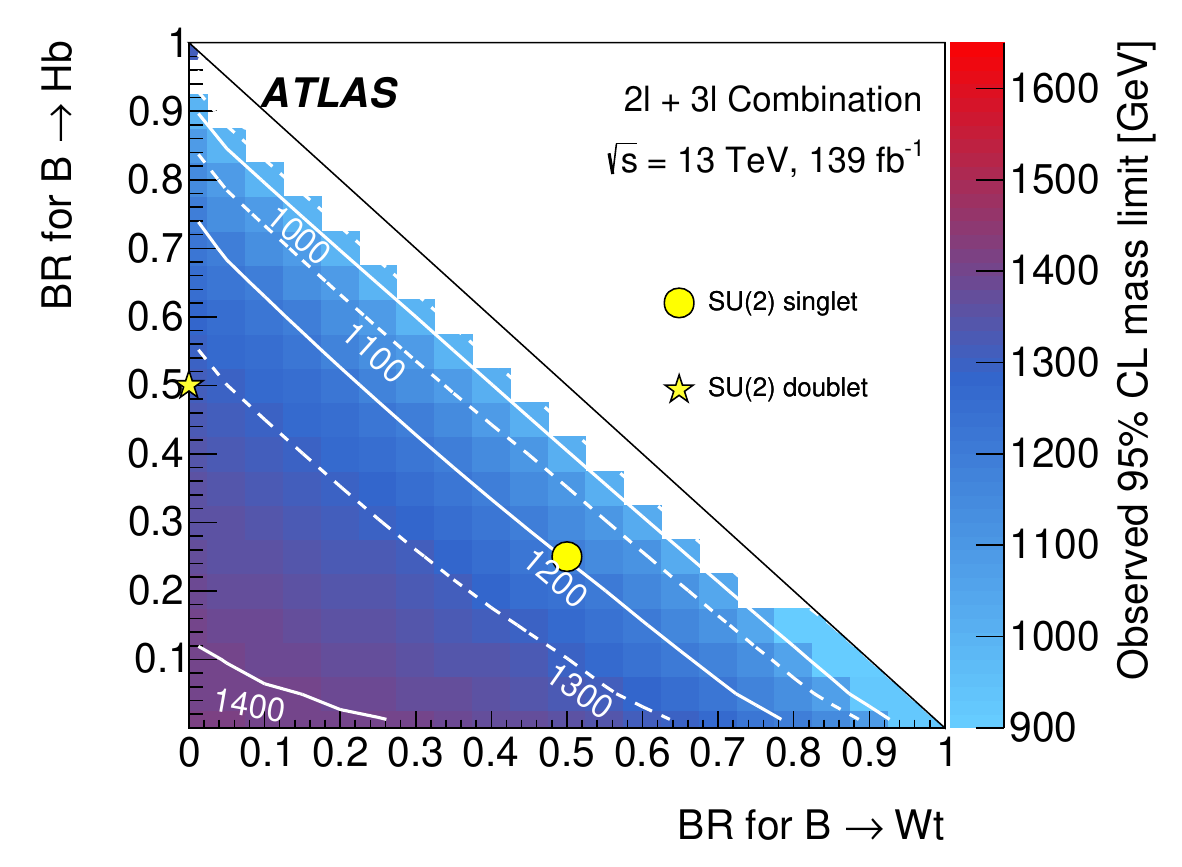}
 
}
 
\caption{The (a,c) expected  and (b,d) observed lower limits on the $T$ and $B$ masses at 95\%~CL
in the \br plane from the combination of the two analysis channels
for the production of (a,b) \TTbar  and (c,d) \BBbar, for
all \br configurations when assuming a total \br of 100\% for
$T\rightarrow Zt$, $T\rightarrow Wb$, and $T\rightarrow Ht$ decays or
$B\rightarrow Zb$, $B\rightarrow Wt$, and $B\rightarrow Hb$ decays,
respectively. The white lines represent the contours of fixed $m_{\mathrm{VLQ}}$.}
\label{fig:ppcomb_limits2d}
\end{figure}

Intermediate \br configurations are obtained by reweighting the simulated signal events
to any target \br composition using generator-level information and assuming
a total \br of 100\% for the $T\rightarrow Zt$, $T\rightarrow Wb$, and $T\rightarrow Ht$ decay modes or the $B\rightarrow Zb$, $B\rightarrow Wt$, and
$B\rightarrow Hb$ decay modes. The reweighting procedure allows 95\% CL upper limits on the production cross-section
to be set across the \br plane for each considered VLQ mass. By comparison with
the theoretical prediction, lower limits on the $T$ and $B$ masses as a function of the \br composition can be obtained, as shown
in \Fig{\ref{fig:ppcomb_limits2d}}. As can be seen,
the analysis is sensitive to a large subset of the possible
\br compositions but the limits
are more stringent closer to the lower-left corner of the
\br plane, and thus for a higher \br for VLQ decays to a $Z$ boson.

\begin{table}[h!]
\centering
\caption{Observed (expected) mass limits in \TeV\ for the $T$ and $B$ singlet,
doublet, and models with 100\% \br to $Z$ for the two
channels and their combination.}
\label{tab:resultMassLimits}
\begin{tabular}{l ccc }
\toprule
\multirow{2}{*}{{Model}} & \multicolumn{3}{c}{{Observed (Expected) Mass Limits [\TeV]}} \\
& {\ppdil} &{\pptril}&{Combination}           \\
\midrule{$T\bar{T}$ Singlet   }          & \dilttsobs\    (\dilttsexp)  &  \trilttsobs\    (\trilttsexp)  & \ttsobs\    (\ttsexp)\\
{$T\bar{T}$ Doublet}             & \dilttdobs\    (\dilttdexp)  &  \trilttdobs\    (\trilttdexp)  & \ttdobs\    (\ttdexp)\\
{$100\%$ $T\rightarrow Zt$}    & \diltthundobs\ (\diltthundexp) &   \triltthundobs\ (\triltthundexp) & \tthundobs\ (\tthundexp)\\
\midrule{$B\bar{B}$ Singlet  }           & \dilbbsobs\    (\dilbbsexp)  & \trilbbsobs\    (\trilbbsexp) & \bbsobs\    (\bbsexp)\\
{$B\bar{B}$ Doublet  }           & \dilbbdobs\    (\dilbbdexp)  & \trilbbdobs\    (\trilbbdexp) & \bbdobs\    (\bbdexp)\\
{$100\%$ $B\rightarrow Zb$ }   & \dilbbhundobs\ (\dilbbhundexp) & \trilbbhundobs\ (\trilbbhundexp)& \bbhundobs\ (\bbhundexp)\\
\bottomrule
\end{tabular}

\end{table}

\FloatBarrier
 
\section{Conclusions}
 
A search for the pair production of vector-like quarks  $T$ or  $B$,
with electric charges $(2/3)e$ and $-(1/3)e$, respectively, is presented
in which at least one of the VLQs decays into a leptonically decaying $Z$ boson
and a third-generation quark.
Two orthogonal channels based on the
number of selected leptons are separately
optimized and their results are statistically combined to obtain the
final result. In both channels a multi-class boosted object tagger for large-radius
jets reclustered from smaller-radius jets is used to
categorize
events according to the numbers of $V$-tags, $H$-tags, and top-tags. The expected SM background is
modelled with MC simulation and a maximum-likelihood fit to the data is performed. No
significant excess over the background
expectation is observed, and therefore 95\% CL upper limits on the $T$ and $B$
pair-production cross-sections are derived.

The combined results exclude $T$ ($B$) masses up to $\SI{\ttsobs}{\TeV}$ and $\SI{\ttdobs}{\TeV}$
(up to $\SI{\bbsobs}{\TeV}$ and $\SI{\bbdobs}{\TeV}$) for the singlet and doublet configurations, respectively.
For the doublet configuration, the lower limits on the $T$ mass are extended by $\SI{90}{\GeV}$ compared to
the most stringent previous ATLAS results,
while the limits on the excluded $T$ and $B$ masses are extended by more than $\SI{200}{\GeV}$
compared to the earlier ATLAS analysis using a subset of the Run~2 data. Finally, for exclusive VLQ decays to
the $Z$ boson and a third-generation quark, this search marginally extends the limits on the $T$ and $B$ masses
to obtain the most stringent limits to date.
 
\section*{Acknowledgements}


We thank CERN for the very successful operation of the LHC, as well as the
support staff from our institutions without whom ATLAS could not be
operated efficiently.
 
We acknowledge the support of
ANPCyT, Argentina;
YerPhI, Armenia;
ARC, Australia;
BMWFW and FWF, Austria;
ANAS, Azerbaijan;
CNPq and FAPESP, Brazil;
NSERC, NRC and CFI, Canada;
CERN;
ANID, Chile;
CAS, MOST and NSFC, China;
Minciencias, Colombia;
MEYS CR, Czech Republic;
DNRF and DNSRC, Denmark;
IN2P3-CNRS and CEA-DRF/IRFU, France;
SRNSFG, Georgia;
BMBF, HGF and MPG, Germany;
GSRI, Greece;
RGC and Hong Kong SAR, China;
ISF and Benoziyo Center, Israel;
INFN, Italy;
MEXT and JSPS, Japan;
CNRST, Morocco;
NWO, Netherlands;
RCN, Norway;
MEiN, Poland;
FCT, Portugal;
MNE/IFA, Romania;
MESTD, Serbia;
MSSR, Slovakia;
ARRS and MIZ\v{S}, Slovenia;
DSI/NRF, South Africa;
MICINN, Spain;
SRC and Wallenberg Foundation, Sweden;
SERI, SNSF and Cantons of Bern and Geneva, Switzerland;
MOST, Taiwan;
TENMAK, T\"urkiye;
STFC, United Kingdom;
DOE and NSF, United States of America.
In addition, individual groups and members have received support from
BCKDF, CANARIE, Compute Canada and CRC, Canada;
PRIMUS 21/SCI/017 and UNCE SCI/013, Czech Republic;
COST, ERC, ERDF, Horizon 2020 and Marie Sk{\l}odowska-Curie Actions, European Union;
Investissements d'Avenir Labex, Investissements d'Avenir Idex and ANR, France;
DFG and AvH Foundation, Germany;
Herakleitos, Thales and Aristeia programmes co-financed by EU-ESF and the Greek NSRF, Greece;
BSF-NSF and MINERVA, Israel;
Norwegian Financial Mechanism 2014-2021, Norway;
NCN and NAWA, Poland;
La Caixa Banking Foundation, CERCA Programme Generalitat de Catalunya and PROMETEO and GenT Programmes Generalitat Valenciana, Spain;
G\"{o}ran Gustafssons Stiftelse, Sweden;
The Royal Society and Leverhulme Trust, United Kingdom.
 
The crucial computing support from all WLCG partners is acknowledged gratefully, in particular from CERN, the ATLAS Tier-1 facilities at TRIUMF (Canada), NDGF (Denmark, Norway, Sweden), CC-IN2P3 (France), KIT/GridKA (Germany), INFN-CNAF (Italy), NL-T1 (Netherlands), PIC (Spain), ASGC (Taiwan), RAL (UK) and BNL (USA), the Tier-2 facilities worldwide and large non-WLCG resource providers. Major contributors of computing resources are listed in Ref.~\cite{ATL-SOFT-PUB-2021-003}.


\printbibliography
 
\clearpage
 
\begin{flushleft}
\hypersetup{urlcolor=black}
{\Large The ATLAS Collaboration}

\bigskip

\AtlasOrcid[0000-0002-6665-4934]{G.~Aad}$^\textrm{\scriptsize 100}$,
\AtlasOrcid[0000-0002-5888-2734]{B.~Abbott}$^\textrm{\scriptsize 117}$,
\AtlasOrcid[0000-0002-7248-3203]{D.C.~Abbott}$^\textrm{\scriptsize 101}$,
\AtlasOrcid[0000-0002-2788-3822]{A.~Abed~Abud}$^\textrm{\scriptsize 34}$,
\AtlasOrcid[0000-0002-1002-1652]{K.~Abeling}$^\textrm{\scriptsize 53}$,
\AtlasOrcid[0000-0002-2987-4006]{D.K.~Abhayasinghe}$^\textrm{\scriptsize 92}$,
\AtlasOrcid[0000-0002-8496-9294]{S.H.~Abidi}$^\textrm{\scriptsize 27}$,
\AtlasOrcid[0000-0001-5329-6640]{H.~Abramowicz}$^\textrm{\scriptsize 149}$,
\AtlasOrcid[0000-0002-1599-2896]{H.~Abreu}$^\textrm{\scriptsize 148}$,
\AtlasOrcid[0000-0003-0403-3697]{Y.~Abulaiti}$^\textrm{\scriptsize 5}$
% ...
\bigskip
\\

% ...
$^{5}$High Energy Physics Division, Argonne National Laboratory, Argonne IL; United States of America.\\
% ...
$^{27}$Physics Department, Brookhaven National Laboratory, Upton NY; United States of America.\\
% ...
$^{34}$CERN, Geneva; Switzerland.\\
% ...
$^{53}$II. Physikalisches Institut, Georg-August-Universit\"{a}t G\"ottingen, G\"ottingen; Germany.\\
% ...
$^{92}$Department of Physics, Royal Holloway University of London, Egham; United Kingdom.\\
% ...
$^{100}$CPPM, Aix-Marseille Universit\'e, CNRS/IN2P3, Marseille; France.\\
$^{101}$Department of Physics, University of Massachusetts, Amherst MA; United States of America.\\
% ...
$^{117}$Homer L. Dodge Department of Physics and Astronomy, University of Oklahoma, Norman OK; United States of America.\\
% ...
$^{148}$Department of Physics, Technion, Israel Institute of Technology, Haifa; Israel.\\
$^{149}$Raymond and Beverly Sackler School of Physics and Astronomy, Tel Aviv University, Tel Aviv; Israel.\\
% ...

\end{flushleft}


\end{document}